\documentclass{article}

\usepackage{arxiv}

\usepackage[utf8]{inputenc} % allow utf-8 input
\usepackage[T1]{fontenc}    % use 8-bit T1 fonts
\usepackage{hyperref}       % hyperlinks
\usepackage{url}            % simple URL typesetting
\usepackage{booktabs}       % professional-quality tables
\usepackage{amsfonts}       % blackboard math symbols
\usepackage{nicefrac}       % compact symbols for 1/2, etc.
\usepackage{microtype}      % microtypography
\usepackage{lipsum,bbm}
\usepackage{graphicx,verbatim,float,subfigure,graphicx,color,siunitx}
\usepackage{amsmath}
\graphicspath{ {./images/} }

\usepackage{amssymb}
\usepackage{svg}
\usepackage{multirow}  %Adds multi-row in tables
\usepackage[export]{adjustbox}
\usepackage{soul}
\usepackage{textcomp}
\usepackage{multirow}
\usepackage{enumitem}

\title{Parameterization-based Neural Network:\\ {\em Predicting Non-linear Stress-Strain Response of Composites}}

\author{
  Haotian Feng\\
  Dept. of Mechanical Engineering \\
  University of Wisconsin-Madison \\
  Madison, WI 53706  \vspace{0.05in} \\
  \And
  Pavana Prabhakar$^*$ \\
  Dept. of Mechanical Engineering \\
  Dept. of Civil \& Env. Engineering \\
  University of Wisconsin-Madison \\
  Madison, WI 53706  \vspace{0.05in} \\
  \texttt{$^*$pavana.prabhakar@wisc.edu}
}

\begin{document}
\maketitle
%\nodate{ }
 
\newcommand{\SPS}[1]{\textcolor{red}{\bf{Sabari: #1}}}
%\newcommand{\VD}[1]{\textcolor{blue}{\bf{ #1}}}
%\newcommand{\pavana}[1]{\textcolor{blue}{\bf{ #1}}}
%----------------------------------------------------------------------------------------
%	TITLE SECTION
%----------------------------------------------------------------------------------------

\begin{abstract}

Composite materials like syntactic foams have complex internal microstructures that manifest high-stress concentrations due to material discontinuities occurring from hollow regions and thin walls of hollow particles or microballoons embedded in a continuous medium. Predicting the mechanical response as non-linear stress-strain curves of such heterogeneous materials from their microstructure is a challenging problem. This is true since various parameters, including the distribution and geometric properties of microballoons, dictate their response to mechanical loading.  To that end, this paper presents a novel Neural Network (NN) framework called Parameterization-based Neural Network (PBNN), where we relate the composite microstructure to the non-linear response through this trained NN model. PBNN represents the stress-strain curve as a parameterized function to reduce the prediction size and predicts the function parameters for different syntactic foam microstructures. We show that compared to several common baseline models considered in this paper, the PBNN can accurately predict non-linear stress-strain responses and the corresponding parameterized functions using smaller datasets. This is enabled by extracting high-level features from the geometry data and tuning the predicted response through an auxiliary term prediction. Although built in the context of the compressive response prediction of syntactic foam composites, our NN framework applies to predict generic non-linear responses for heterogeneous materials with internal microstructures. Hence, our novel PBNN is anticipated to inspire more parameterization-related studies in different Machine Learning methods.

%Predicting non-linear stress-strain responses of heterogeneous materials from their microstructure is a challenging problem, especially for composite materials like syntactic foam composites. Syntactic foam composites manifest high-stress concentrations due to material discontinuities occurring from hollow regions and thin walls of the micro balloons embedded in a continuous medium. Various parameters, including the distribution and geometric properties of microballoons, affect the foam’s mechanical responses, that is, stress-strain curves. To that end, this paper presents a novel \textbf{Parameterization-based Neural Network (PBNN)} to predict the non-linear stress-strain responses of foam composites from their microstructures. PBNN represents the stress-strain curve as a parameterized function to reduce the size of prediction and predicts the function parameters for different syntactic foam microstructures. Results show that this approach enables more accurate non-linear stress-strain response predictions and corresponding parameterized functions using smaller datasets than existing approaches. The PBNN utilizes a Feature Extraction module to extract high-level features from the geometry data and incorporates a Modification module to tune the predicted response through an auxiliary term prediction. The novel PBNN is significant for facilitating non-linear curve predictions for heterogeneous materials, like composites, and is anticipated to inspire more parameterization-related studies in different Machine Learning methods.

\end{abstract}

% keywords can be removed
\keywords{Parameterized Function \and Neural Network \and Stress-Strain Curve \and Reinforced Composites \and Finite Element Analysis}

%----------------------------------------------------------------------------------------
%	ARTICLE CONTENTS
%----------------------------------------------------------------------------------------

\section{Introduction}\label{intro}

This paper presents a novel Neural Network (NN) for predicting the non-linear stress-strain behavior of heterogeneous materials through a parameterization-based approach. Specifically, we relate the composite microstructure to the non-linear response through this trained NN model. We build this framework in the context of the compressive response prediction of polymeric foams called `syntactic foams' with their microstructure as the inputs. Syntactic foams have complex internal microstructures with high-stress concentration regions and macroscale non-linear compressive stress-strain responses, making it more challenging to predict their stress-strain responses. To that end, the NN presented in this paper applies to predicting generic non-linear responses for composites with internal microstructures.

Syntactic foams, specifically polymer ones, are closed-cell composite foams with hollow spheres or particles called `microballoons' embedded in a polymer matrix. The presence of hollow spheres results in several excellent mechanical properties, including lower density, higher specific strength, lower thermal expansion coefficient, and lower moisture absorption\cite{gupta2004microballoon,gladysz2006syntactic}. Due to these properties, syntactic foam materials are widely used as buoyancy materials for marine applications as a component of sea-related products and offshore products\cite{choqueuse2008ageing,gupta2014applications}. Moreover, syntactic foams have been extended to other applications like the aerospace and automotive industry. For effective design and optimization, such extended applications require predicting their mechanical properties, especially the stress-strain responses.

Past researchers have focused on establishing the mechanical properties of syntactic foams through experimental testing and numerical studies. Gupta et al.\cite{gupta2004compression} performed compression tests and showed that compressive strength and modulus would increase when the microballoon radius decreases. Shahapurkar et al.\cite{shahapurkar2018compressive} experimentally showed that compressive strength decreased with increasing cenosphere volume fraction for both modified and unmodified surfaces. Moreover, the authors showed a reduction in the compressive modulus but an increase in the compressive strength for arctic-conditioned samples. Jayavardhan et al.\cite{jayavardhan2018quasi} conducted quasi-static compressive testing of glass microballoon reinforced high-density polyethylene syntactic foams with different densities. The authors showed that the compressive modulus increased as the microballoon volume fraction increased, but the yield strength, densification stress, and overall energy absorption were reduced. Exploring the influence of many parameters of syntactic foams experimentally is challenging and time-consuming. Thus later, Prabhakar et al.\cite{prabhakar2022densification} developed the computational modeling to establish a fundamental understanding of densification mechanics of polymeric syntactic foams under compressive loading accounting for microballoon volume fraction, microballoon wall thickness, bonding between the microballoons and the matrix, and the crushing strength of microballoons. The authors further utilized multiple linear regression to understand the influence of structural and material parameters on its densification properties. Wang et al.\cite{wang2022influence} investigated how the strength distribution of a batch of hollow glass microspheres (HGM) influences the compression strength of syntactic foams. The authors discovered that the compressive strength of syntactic foams improved with an increase in HGM's strength.

The above research explored the syntactic foam composite's time-independent (\textbf{quasi-static}) mechanical properties. However, analyzing how mechanical properties of syntactic foam composite change under different strains or loading rates is also essential, i.e., the strain-dependent (\textbf{dynamic}) mechanical properties. Woldesenbet et al.\cite{woldesenbet2005effects} analyzed the effect of density and strain rate on the properties of syntactic foam and showed a considerable increase in peak strength of syntactic foams for higher strain rates and higher density. Song et al.\cite{song2005confinement} investigated the dynamic compressive properties of an epoxy syntactic foam at various strain rates under lateral confinement with a split Hopkinson pressure bar. The authors discovered that the quasi-static and dynamic stress-strain behavior has an elastic-plastic-like shape, whereas an elastic-brittle behavior was observed under uniaxial loading. Li et al.\cite{li2009strain} experimentally analyzed the compressive responses of glass microballoon epoxy syntactic foams within a range of strain rates. The authors combined testing results with finite element stress analysis to determine the foam's localized damage and failure modes. Shunmugasamy et al.\cite{shunmugasamy2010strain} utilized microCT-scanning and scanning electron microscopy to understand the effect of high strain rate loading on the deformation and fracture characteristics of syntactic foams and further understand the strain rate dependence of failure mechanisms. Zhang et al.\cite{zhang2022dependency} investigated the compressive response of epoxy syntactic foam with strain rate and temperature dependency. The authors also developed a non-linear phenomenological model to describe the responses of syntactic foam and its temperature-strain rate equivalence. These works have shown the different approaches (experimental and analytical) to obtaining the strain-dependent mechanical properties of syntactic foam composites. However, these analyses only focus on limited microballoon distributions by considering only a few samples. To explore the full range of parameters of syntactic foams and to design these foams effectively, an approach to relate the stress-strain responses to microballoon properties and distributions is needed. To achieve this, we will utilize Deep Learning methods to determine the complete stress-strain responses of syntactic foams given the volume fraction and wall thickness of randomly distributed microballoons.

The emergence of Machine Learning methods facilitates research to predict the mechanical properties of composite materials with the graph neural network as an essential tool. Graph Neural Network is developed based on Deep Convolutional Neural Network\cite{krizhevsky2012imagenet} (DCNN) and Generative Adversarial Network\cite{creswell2018generative} (GAN). Researchers have been focusing on solving engineering design and analysis problems with Graph Neural Networks, like~\cite{zhou2022study,jiang2021stressgan}. Regarding composite materials, Chen et al.\cite{chen2019machine} compared how different Machine Learning techniques, like regression model, DCNN, and Gaussian process, can accelerate the composite material design. Feng et al.\cite{feng2021difference} proposed a Difference-based Neural Network to enhance the stress distribution prediction within different composite micromechanical models, especially for models with stress concentrations in the stress distribution contours. Sepasdar et al.\cite{sepasdar2021data} proposed a modified U-Net framework to predict the damage and failure within microstructure-dependent composite materials. Feng et al.\cite{feng2022physics} further proposed a Physics-Constraint Neural Network to understand the forward and inverse predictions of woven composite models in the mesoscale. 

Besides predicting the linear elastic mechanical properties with Machine Learning, researchers also utilized Machine Learning to predict the non-linear constitutive behaviors and mechanical responses of composite materials, like the entire stress-strain curve. Hashash et al.\cite{hashash2004numerical} developed a Neural Network constitutive model to replace the commonly used integration procedures in Finite Element Analysis. Then a consistent material stiffness matrix is derived based on the Neural Network constitutive model instead of conventional plasticity-based models. This new model leads to efficient convergence of the Finite Element Newton iterations. Bos et al.\cite{du2020modeling} developed a Neural Network-based constitutive model to capture the elastic-plastic stress-strain curve. The authors proposed to sub-sample the stress-strain curve at several discrete points and then predicted the stress values at different discrete strain values. The final curve was obtained by interpolating the discrete points. However, this approach still needs several discrete points for prediction, and there are errors during interpolation. Yang et al.\cite{yang2020prediction} combined principal component analysis (PCA) and convolutional neural networks to predict the entire stress-strain behavior of binary composite materials. The authors showed that the PCA could effectively transform the stress-strain curve into a latent space and the prediction error is less than 10\%. Kosmerl et al.\cite{kovsmerl2022predicting} proposed a Neural Network by combining a convolutional neural network and residual neural network to predict the stress-strain curve for single-walled carbon nanotube configurations. These methods have shown the promising aspect of Neural Networks in obtaining the non-linear stress-strain behaviors of a targeting model. However, one key drawback of the above methods is that we need a large training dataset to better predict and represent the latent space features. Such massive training datasets are usually extremely challenging to generate as they come from experiments or numerical simulations. 

In this paper, we propose a \textbf{Parameterization-based Neural Network (PBNN)} where we represent the non-linear stress-strain response of the chosen composite (here hollow particle reinforced geometries) with a parameterized function space. The parameters in the function can be different for different stress-strain curves. Then, we utilize the concepts of self-supervised learning\cite{jing2020self} and transfer learning\cite{weiss2016survey} to effectively extract the latent features from the composite geometric model using an Encoder-Decoder Neural Network. The key benefits of our approach are:

\begin{enumerate}
    \item We use an Encoder-Decoder Neural Network for latent feature extraction from composite geometries (named Feature Extraction module). This Feature Extraction module reduces the input dimension from a 256-by-256 image to a 128-by-1 vector, simplifying the following Neural Network prediction task. It is easier to generate different matrices to represent the composite geometries, while it is time-consuming to solve each model numerically.
    
    \item We use a parameterized representation of the stress-strain curve to capture the shape of the true stress-strain response. Otherwise, we usually need a `physically meaningful' or higher-order smooth function to represent the stress-strain curve. This parameterized representation vastly reduces the prediction size as it requires fewer data during training to achieve a relatively good prediction.

    \item We propose a module named `Modification module' that predicts an auxiliary term to increase the stress-strain curve prediction accuracy. This Modification module serves as a constraint by predicting an extra data point on the stress-strain curve. Then it modifies the stress-strain curve predicted through Neural Network by adding a carefully constructed polynomial equation.
    
    \item Our proposed PBNN is not limited to the stress-strain curve prediction with a known function but works for other response prediction problems. For example, we can fit any arbitrary curve with a polynomial function, which will be the targeting parameterized function for our proposed PBNN.
\end{enumerate}

\section{Overview of the Machine Learning Framework} \label{sec:overview}

\begin{figure}[h!]
\centering
	\includegraphics[width=0.95\textwidth]{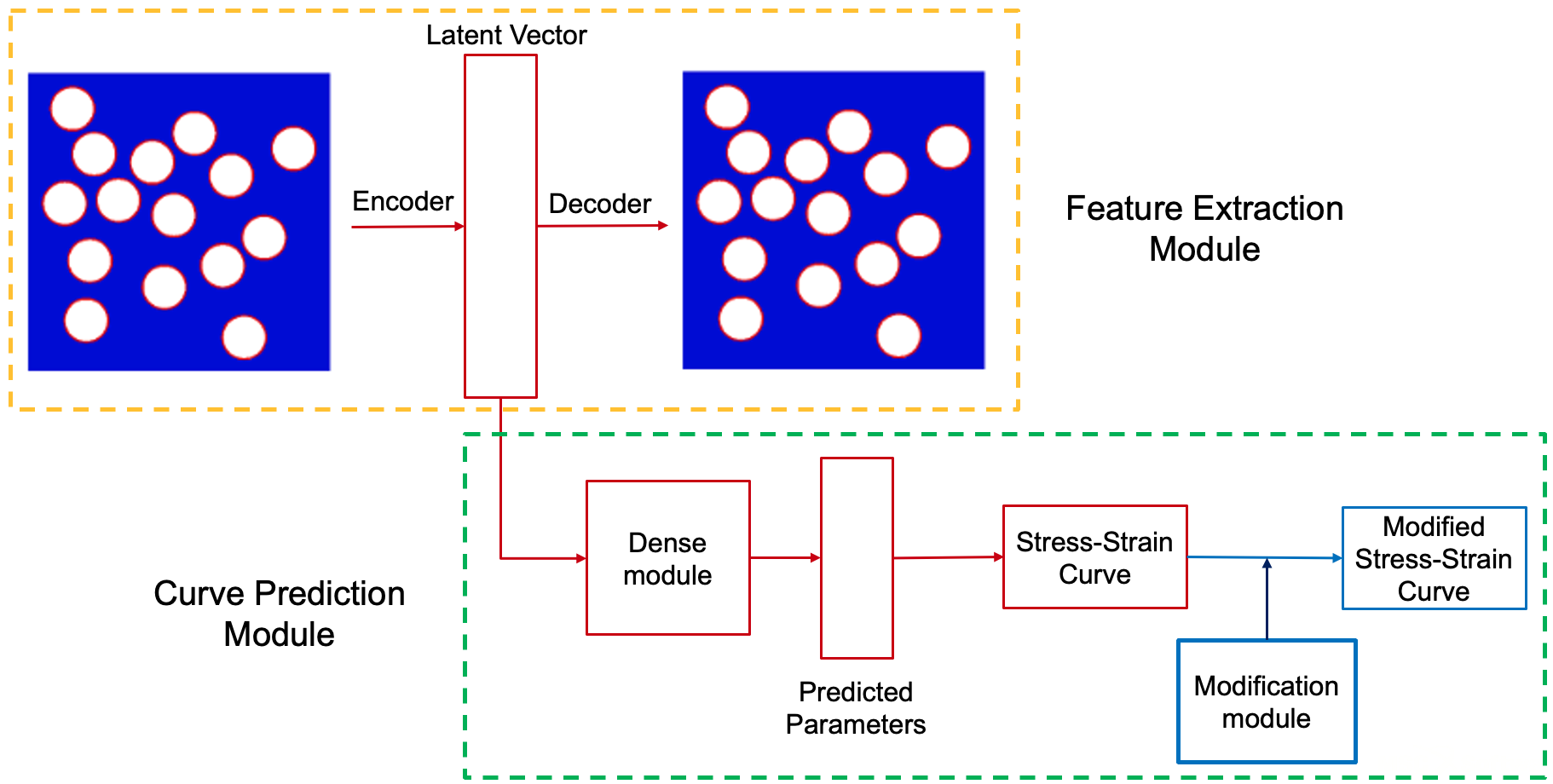}
	\caption{Overall framework for predicting the non-linear mechanical responses with PBNN: (1) Orange block is the Feature Extraction module, which extracts the high-level features (Latent Vector) of the syntactic foam geometry using an Encoder-Decoder structure. (2) Green block is the Curve Prediction module, which predicts the stress-strain curve from the extracted high-level features using a Dense module and a Modification module.}
\label{img:overall}
\end{figure}

The PBNN framework consists of two key modules: 1) the Feature Extraction module and 2) the Curve Prediction module, as shown in Figure~\ref{img:overall}. The Feature Extraction module extracts a high-level feature vector from the syntactic foam geometry. Then the feature vector is brought into the Curve Prediction module, consisting mainly of a Dense module and a Modification module, to predict the final stress-strain curve. We perform the Machine Learning training on NVIDIA GeForce RTX 2080 SUPER with 3072 CUDA cores. 
\begin{comment}
   We have provided access to our implemented Machine Learning framework code on our GitHub page and training data on Google Drive, as mentioned in the `Data Availability'~\ref{dataAvail} section at the end of this paper. 
\end{comment}

\subsection{Feature Extraction Module}\label{sec:feature_extraction}
\begin{figure}[h!]
\centering
	\includegraphics[width=0.95\textwidth]{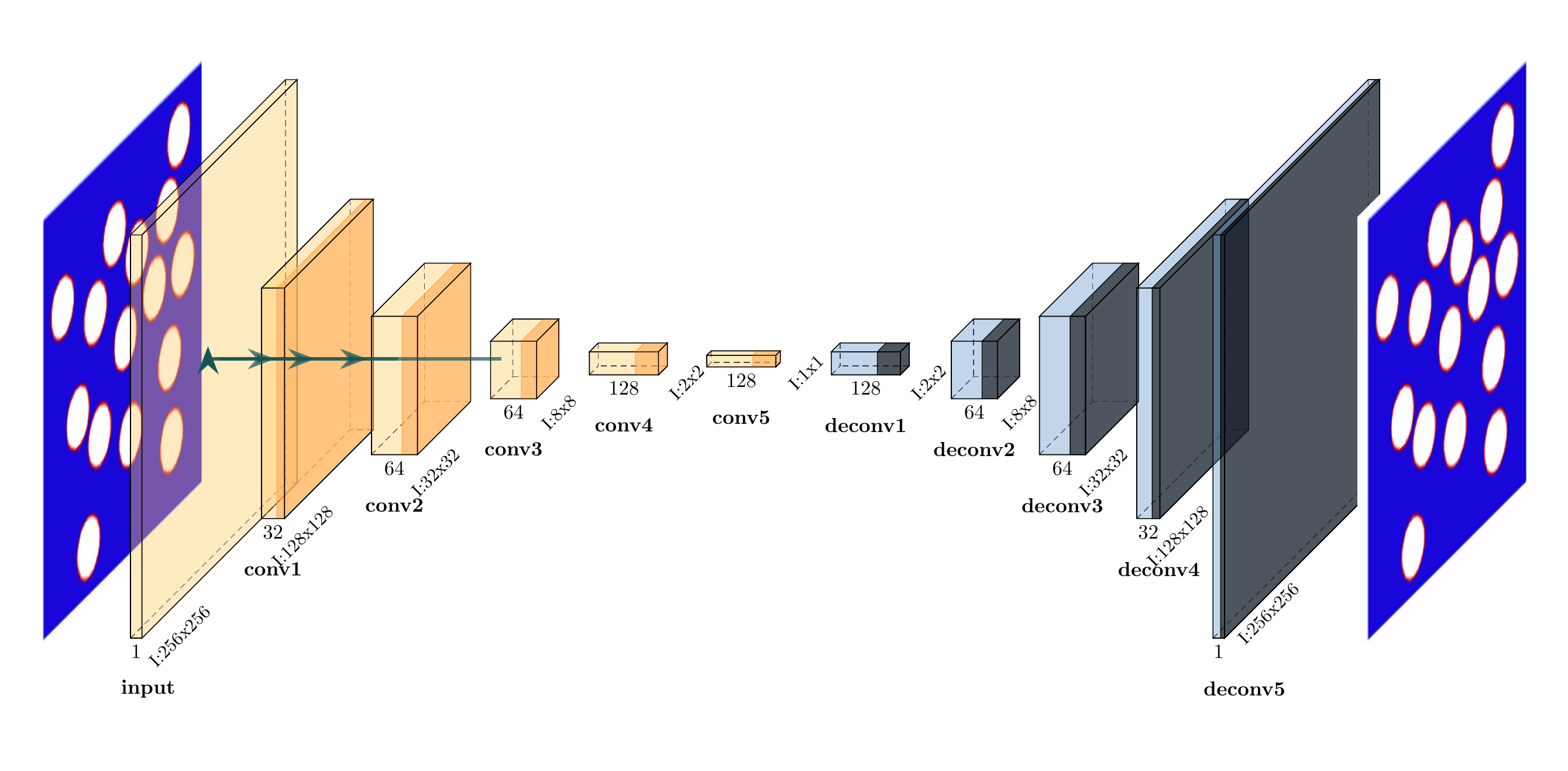}
	\caption{Feature Extraction module framework: the yellow color blocks belong to the Encoder module, and the blue color blocks belong to the Decoder module, as shown in Figure~\ref{img:overall}. Light yellow and blue blocks are convolutional and deconvolutional layers, respectively. The dark yellow and dark blue colors are the ReLU activation layer. Block `conv5' outputs a `Latent Vector,' the high-level feature vector of the syntactic foam geometry. It will be further used in the Curve Prediction module.}
\label{img:Feature_Extraction}
\end{figure}

The framework for the Feature Extraction module is shown in Figure~\ref{img:Feature_Extraction}, which is constructed based on the Encoder-Decoder structure. The Encoder will extract the high-level features from the input model, and the Decoder will expand the extracted high-level features back to the input model. Here the Decoder is needed to train the Feature Extraction Module, as the loss function of the Feature Extraction Module is defined based on the input model. The module simplifies the Neural Network training problem by reducing the complex 256-by-256 image input into a 128-by-1 vector. This feature vector can be treated as a high-level equivalence of the corresponding input geometry. Then the prediction is simplified into a problem as predicting the stress-strain curve from a 128-by-1 feature vector instead of a 256-by-256 image matrix. 

\subsection{Curve Prediction module and Modification module}\label{sec:curve_pred_module}

The Curve Prediction module predicts the stress-strain curve from the extracted feature vector. This module consists of two main sub-modules: the Dense and Modification modules.
The Dense module consists of three dense layers. The first two dense layers have 64 neurons at each layer, with a tanh activation function. A linear activation function follows the last dense layer and has a neuron size equal to the prediction size, which equals the size of function parameters plus one. For example, if the size of targeting parameters is three, the last dense layer will predict four neurons; three are function parameters, and the last is the `end stress'. The `end stress' is an auxiliary term that refers to the stress value at the maximum strain considered (15\% strain in this paper). Predicted function parameters will determine the initial expression of the stress-strain curve. The Modification module will utilize the predicted `end stress' to form a `Modification function', which will be further added to the initial expression to reconstruct the initial curve by shifting it closer to the true stress-strain curve. 

The Modification module is a critical component of this prediction framework. By predicting the function parameters, the value of each point on the curve will be a non-linear function of all coefficients, making it extremely challenging to add constraints to the values of these points. Moreover, the shape of the function can be sensitive to given parameters, so adjusting the function parameters might significantly change the curve shape. Thus, we need to add additional constraints to the predicted curve but simultaneously keep the prediction physically meaningful. So we incorporated the Modification function to serve as the constraint. The Modification function can be expressed as Equation~\ref{eqn:modify_func}.

\begin{equation}
    g(\lambda)=(\sigma_{end}-P(\lambda=1.15))(\lambda-1)^2/0.15^2
\label{eqn:modify_func}
\end{equation}

Where $\lambda$ is the principle stretch defined as the ratio of the deformed length to the undeformed length along the principal axes. When a uniaxial stress is applied, $\lambda$ is related to uniaxial strain $\epsilon$ as $\lambda=\epsilon+1$. $\sigma_{end}$ is the auxiliary prediction term, denoting the predicted `end stress' from the dense module corresponding to stretch $\lambda$=1.15. $P(\lambda)$ denotes the fitting function we choose to represent the stress-strain curve. Detailed expression of function $P(\lambda)$ considered in this paper can be referred to in Section~\ref{sec:ss_curve_rep}. $P(\lambda=1.15)$ denotes the true value of `end stress' obtained from the fitting function. The term $P(\lambda=1.15)$ ensures the `end stress' of the final stress-strain curve matches the predicted `end stress'. `0.15' is the normalization term, representing the maximum strain for the stress-strain response considered. This normalization term ensures that the $g(\lambda=1.15)$ represents the gap between true `end stress' and predicted `end stress'. This $g(\lambda=1.15)$ is effectively pulling the `end stress' of the predicted curve to the `end stress' of the true curve. Since this normalization term depends on the maximum strain considered, it needs to be updated accordingly for a different dataset used. We designed the Modification function as a quadratic function because the stress prediction error generally increases as the strain value increases and is easier to construct. The base term $\lambda-1$ ensures the initial point of the curve is not shifted after adding the Modification function. When $\lambda$=1, there is no change to the final stress-strain curve. At the `end stress', when $\lambda$=1.15, we only end up with $\sigma_{end}-P(\lambda=1.15)$. Then the final stress-strain curve expression will be $F(\lambda)=P(\lambda)+g(\lambda)$. 

%An example of Curve Prediction Structure on the Ogden function prediction can be referred to Figure~\ref{img:pbnn_mod}(b).

%\vspace{-0.2in}
\section{Syntactic Foam Computational Modelling} \label{se:modelDefinition}

The syntactic foam micromechanical models are modeled using the Finite Element Method. A description of computational modeling, including boundary conditions and materials, is presented in Section~\ref{se:bvp} and~\ref{sec:material}. The parametric space of syntactic foam models and corresponding stress-strain curves are shown in Section~\ref{sec:para_space}.

\subsection{Modelling and Boundary Value Problem}\label{se:bvp}
This paper considers 2D micromechanical syntactic foam models with randomly distributed microballoons in a matrix region. All micromechanical model dimensions are maintained at 0.295 mm x 0.295 mm, and the microballoon outer radius is maintained at 0.0225 mm. These micromechanical models are implemented within Finite Element software ABAQUS\cite{abaqus} with a mesh consisting of linear plane stress elements (CPS3 and CPS4). The mesh size is determined through mesh convergence analysis such that results are within 10\% of the converged solution to reach a balance between convergence and computational cost. Since we have not considered strain-softening material behavior in our models, there is no pathological mesh dependency. 

\begin{figure}[h!]
\centering
\includegraphics[width=0.5\textwidth]{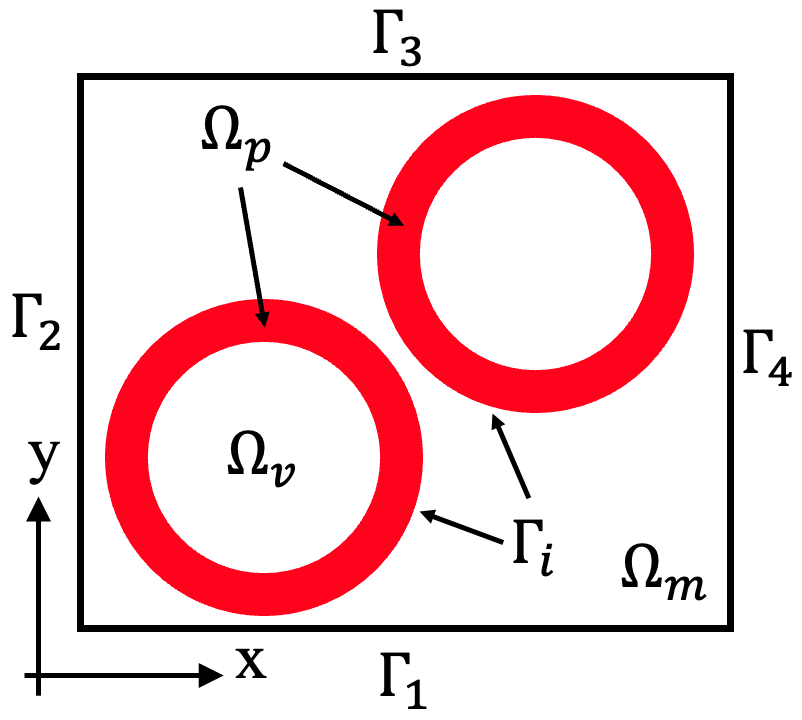}
\caption{An example of 2D syntactic foam micromechanical model}
\label{img:2d_syntac_foam}
\end{figure}

A schematic representation of the syntactic foam micromechanical model is shown in Figure~\hbox{\ref{img:2d_syntac_foam}}. Here, $\Omega_{m}$ represents the matrix region, $\Omega_{p}$ represents the microballoon wall region, and $\Omega_{v}$ represents the hollow region or void inside the microballoons. $\Gamma_{1-4}$ are the external boundaries of the micromechanics domain, and $\Gamma_{i}$ are the interfaces between each microballoon and the matrix region. The volume fraction of matrix is given by V$_{m}=\frac{\Omega_{m}}{\Omega}$. The volume fraction of microballoons, including that of the particle wall and void, in a syntactic foam composite, is V$_{mb}=1-$V$_{m} = \frac{\Omega_{p}+\Omega_{v}}{\Omega}$. 

Compressive displacement $\Delta_{applied}$ is applied on $\Gamma_{4}$. $\Gamma_{1}$ is fixed from deforming only along the y-axis and $\Gamma_{2}$ only along the x-axis. A flat boundary condition is considered on $\Gamma_{3}$ to achieve uniform deformation along the y-axis. These boundary conditions are commonly used in micromechanics modeling of composites, and the corresponding boundary conditions are:

%\vspace{-0.6in}
\begin{equation}\label{appDisp}
\begin{split}
    \hat{u}_{x} = - \Delta_{applied} \quad \textrm{on} \quad \Gamma_{4} \\
 \hat{u}_{x} = 0 \quad \textrm{on} \quad \Gamma_{2} \\
 \hat{u}_{y} = 0 \quad \textrm{on} \quad \Gamma_{1} \\
 \hat{u}_{y} = \textrm{constant} \; \textrm{on} \; \Gamma_{3}
 \end{split}
\end{equation}

\subsection{Constitutive Materials Description}\label{sec:material}
This paper considers High-Density Polyethylene (HDPE) as the material for the matrix. HDPE manifests a non-linear behavior under compression. The compressive stress-strain material data for HDPE is obtained from \hbox{\cite{jayavardhan2018quasi,prabhakar2022densification}}, and is used as the input to fit an Ogden hyperelastic model. Prabhakar et al.\hbox{\cite{prabhakar2022densification}} has shown that the Ogden function effectively represents the stress-strain relationship for pure HDPE material. The glass microballoons (GMBs) are modeled as a linear elastic material. HDPE and GMB material properties can be found respectively in \hbox{\cite{prabhakar2022densification}}. This study considers the interfacial behavior between the GMBs and HDPE matrix as perfectly bonded.

\begin{comment}
The fitted parameters of Ogden function is shown in Table~\ref{tab:ogdenFit}.

\begin{table}[h!]
\caption{Parameters of the fitted Ogden model}
\centering
\resizebox{0.6\textwidth}{!}{
\begin{tabular}{cccc}    
\hline 
i  & $\mu_i$ (MPa)& $\alpha_i$ & D$_i$ \\ \hline 
1  &  14.9637655 &  1.14899635 &  2.603E-03 \\ 
2  &    212.735555    &    11.8906340 & 0  \\   
3  & -106.383163  &    -5.94506580 & 0 \\  \hline
\end{tabular}
}
\label{tab:ogdenFit}
\end{table}
\end{comment}

\subsection{Parametric Space and Stress-strain Curves}\label{sec:para_space}
\begin{figure}[h!]
\centering
\subfigure[]{
  \includegraphics[width=0.3\textwidth,height=0.3\textwidth]{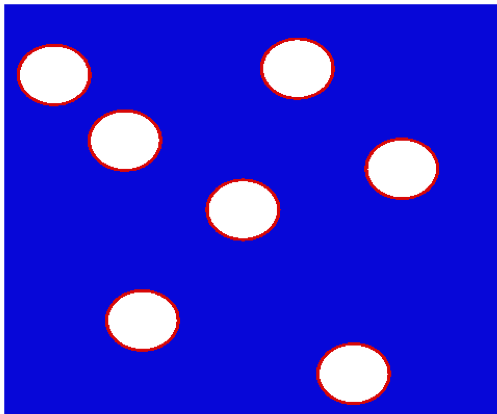}
}
\centering
\subfigure[]{
  \includegraphics[width=0.3\textwidth,height=0.3\textwidth]{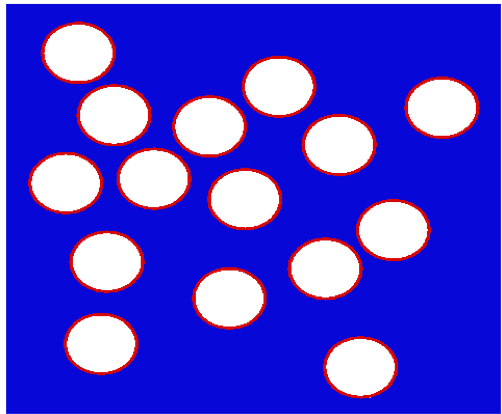}
}
\centering
\subfigure[]{
  \includegraphics[width=0.3\textwidth,height=0.3\textwidth]{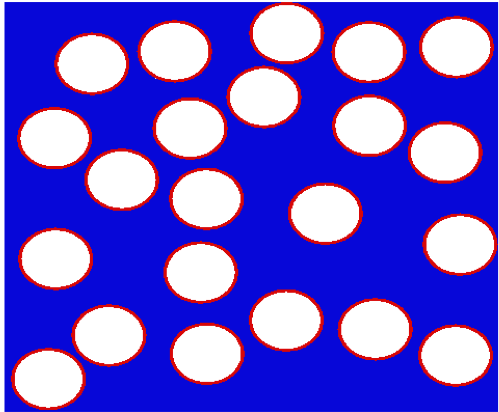}
}
\caption{Example of 2D RUC models of thick-wall syntactic foam geometries with different volume fractions of (a) 10\% (b) 30\% (c) 50\%: Blue color refers to the matrix, red color refers to the wall of microballoon, and white color refers to the hollow region.}
\label{img:syntac_model}
\end{figure}

The syntactic foam micromechanical model has several significant geometric parameters which govern the parametric space: (1) Number of microballoons: We consider the microballoon volume fraction to range from 10\% to 50\%. The volume fraction represents the ratio between the volume occupied by the microballoon and the whole model. A higher volume fraction implies more microballoons within a fixed matrix volume. (2) Position of microballoons: Each microballoon can be randomly located within the matrix. 
%Different distributions will largely affect the stress-strain response of syntactic foam models.
(3) Size of microballoons: We consider uniform microballoon outer diameters, but two different microballoon wall thicknesses are considered: 1.08 and 2.16 micrometers. The 1.08-micrometer thickness is denoted as `thin-wall', and the 2.16-micrometer thickness is denoted as `thick-wall'. The wall-thickness significantly affects the mechanical properties of syntactic foam composites, including stress-strain response, modulus, and failure modes \cite{gupta2004microballoon}. Examples of different syntactic foam micromechanical models are shown in Figure~\ref{img:syntac_model}.

\begin{figure}[h!]
\centering
\subfigure[]{
  \includegraphics[width=0.45\textwidth,height=0.3\textwidth]{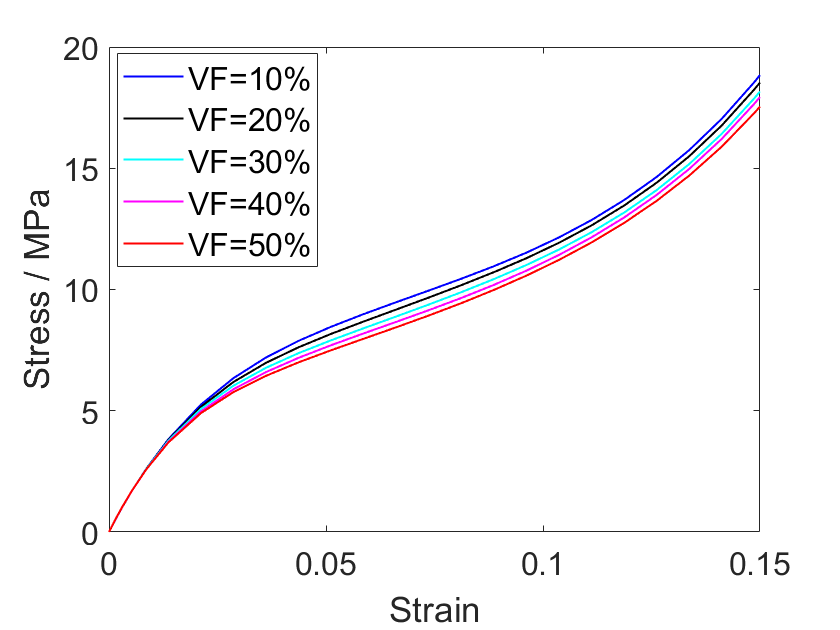}
}
\centering
\subfigure[]{
  \includegraphics[width=0.45\textwidth,height=0.3\textwidth]{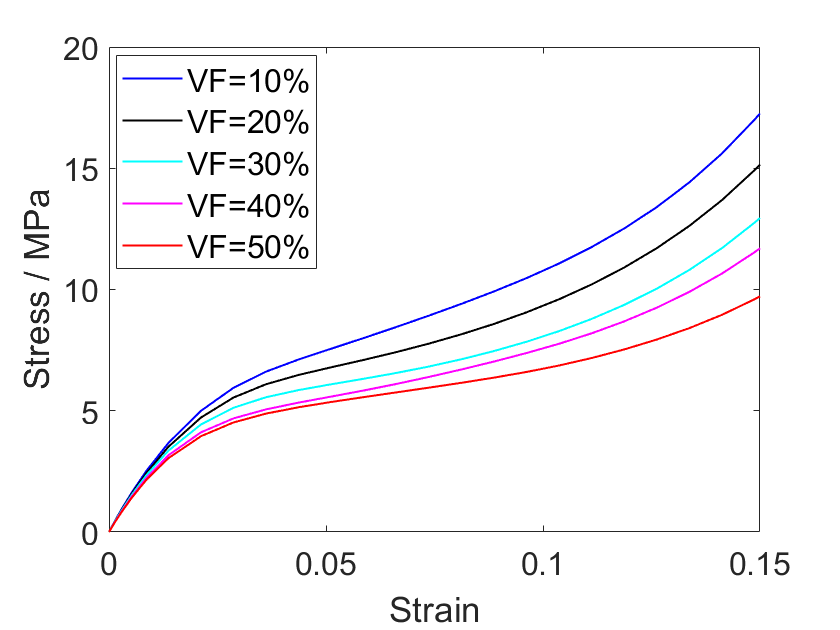}
}
\caption{Example of thick-wall and thin-wall syntactic foam stress-strain curves with different volume fractions: (a) thick-wall syntactic foam stress-strain curves; (b) thin-wall syntactic foam stress-strain curves. `VF=10\%' denotes 10\% microballoon volume fraction}
\label{img:syntac_ss}
\end{figure}

This paper solves 6825 thin-wall and thick-wall syntactic foam models to obtain their stress-strain curves. Figure~\ref{img:syntac_ss} shows sample stress-strain curves of thick-wall and thin-wall syntactic foam models with different microballoons volume fractions. 

\section{Machine Learning Model Inputs}\label{se:compModel}

Two inputs are used to set up the Machine Learning training process: the syntactic foam geometry and the corresponding stress-strain curve. We use 18,000 syntactic foam geometries, where 6825 models are solved to obtain their stress-strain responses. That is, 18,000 syntactic foam geometries were used to train the Feature Extraction module, and the 6825 syntactic foam models (thick-wall and thin-wall) were used to obtain stress-strain curves for further training the Curve Prediction module. 

\subsection{Syntactic Foam Geometry Representation}\label{sec:geom_represent}

The 2D geometric representation of a syntactic foam includes three parts: the matrix labeled as `1', the microballoon wall labeled as `2', and the hollow region labeled as `0' shown in Figure~\ref{img:syntac_model} in blue, red, and white, respectively. Since the wall of the microballoon is very thin compared to the whole model, we construct a 256-by-256 Cartesian Map to ensure at least three layers of pixels along the microballoon wall thickness. Since boundary conditions are consistent for all models to calculate the stress-strain curve, boundary conditions are not considered part of the inputs.

\subsection{Stress-Strain Curve Representation} \label{sec:ss_curve_rep}
The stress-strain curves for the syntactic foams are non-linear responses relating the compressive stresses with strains on the micromechanical RUC models. To predict the stress-strain curves using Machine Learning algorithms, we can treat them as a piecewise function by connecting several discrete points or a smooth function with a particular expression. In this paper, we consider three different representations: (1) linear piecewise function, (2) cubic polynomial function, and (3) Ogden function. We utilize the Mean Squared Error (MSE), defined as Equation~\ref{eqn:mse_func}, to quantify the fitting errors of different representations.

\begin{equation}
    MSE = \frac{1}{n}\sum_{i=1}^n (Y_i - \hat{Y}_i)^2
\label{eqn:mse_func}
\end{equation}

\noindent where n is the number of data points, $Y_i$ is the i-th observed value and $\hat{Y}_i$ is the i-th predicted value.

\subsubsection{Linear piecewise function representation}
To balance the need between capturing the shape of the curve with a linear piecewise function and not making the Machine Learning prediction too difficult, we consider 21 uniformly distributed discrete strain values from 0 to 15\%, corresponding to 21 discrete points along the stress-strain curve. Then the PBNN will attempt to directly predict these 21 discrete points to represent the stress-strain response. 
\begin{comment}
The values at each discrete point are calculated using linear interpolation. For a domain with two fixed endpoints $(x_0,y_0)$ and $(x_1,y_1)$ for strain and stress combinations, the corresponding value $y$ at each discrete point $x$ is given by Equation~\ref{eqn:linear_interp}.

\begin{equation}
    y = y_0 + \frac{y_1-y_0}{x_1-x_0}(x-x_0)
\label{eqn:linear_interp}
\end{equation}
\end{comment}
After obtaining the values for all 21 discrete points, the overall curve will be the union of several discrete linear piecewise functions in different subdomains. Each j-th subdomain consists of two points $X_j$ and $X_{j+1}$, where $X_j$ is the j-th discrete strain point having the corresponding stress value of $Y_j$. The linear piecewise function in the j-th subdomain can be expressed as in Equation~\ref{eqn:piece_linear}.

\begin{equation}
    Y = F_j(X) = Y_j + \frac{Y_{j+1}-Y_j}{X_{j+1}-X_j}(X-X_j)
\label{eqn:piece_linear}
\end{equation}

where $X\in[X_j,X_{j+1}]$. The ultimate linear piecewise function $P(\lambda)$ can be expressed as $P(\lambda)=\bigcup_{j} F_j(\lambda) \,\, for \,\, \forall \lambda\in[X_j, X_{j+1}]$, $\lambda=\epsilon+1$ is the principle stretch value in the defined domain of interest, as mentioned in Section~\ref{sec:curve_pred_module} and$\epsilon$ refers to the strain value. $j$ is a collection of all subdomains.

\subsubsection{Cubic polynomial function representation}
A polynomial function can also be used to represent the stress-strain curve. By testing the fitting errors of polynomials with different order using polynomial regression\cite{harrell1984regression}, we pick the cubic polynomial function expression and calculate the fitting coefficients. Since we know the stress value is zero when strain is zero or $\lambda$ is 1, we carefully constructed the cubic polynomial function as Equation~\ref{eqn:cubic_func}.
\begin{equation}
    P(\lambda) = a_3(\lambda-1)^3 + a_2(\lambda-1)^2 + a_1(\lambda-1)
\label{eqn:cubic_func}
\end{equation}

\begin{figure}[h!]
\centering
\subfigure[]{
  \includegraphics[width=0.3\textwidth,height=0.23\textwidth]{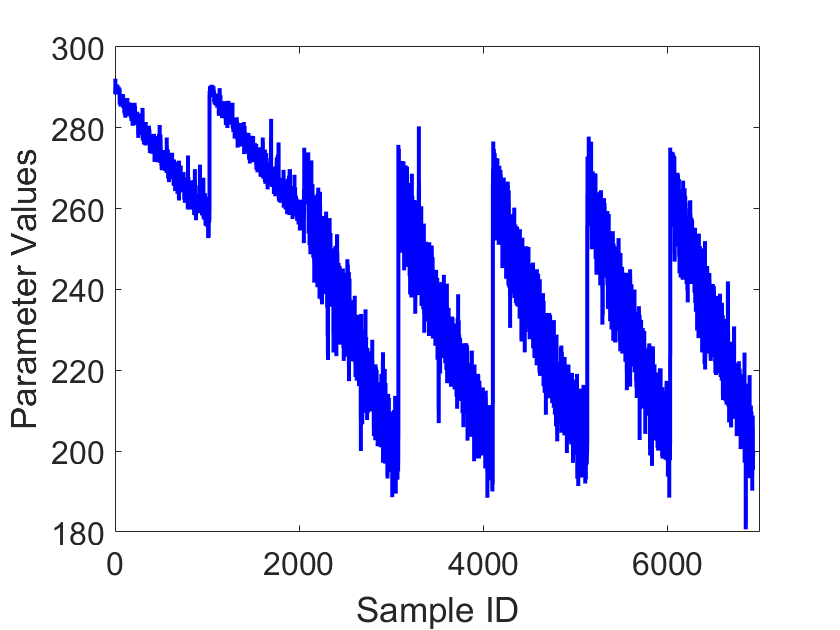}
}
\centering
\subfigure[]{
  \includegraphics[width=0.3\textwidth,height=0.23\textwidth]{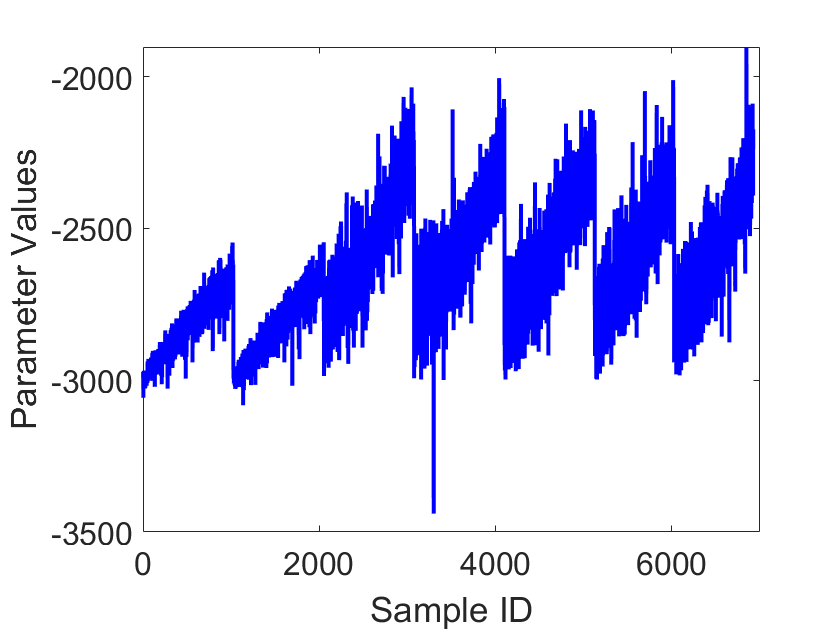}
}
\centering
\subfigure[]{
  \includegraphics[width=0.3\textwidth,height=0.23\textwidth]{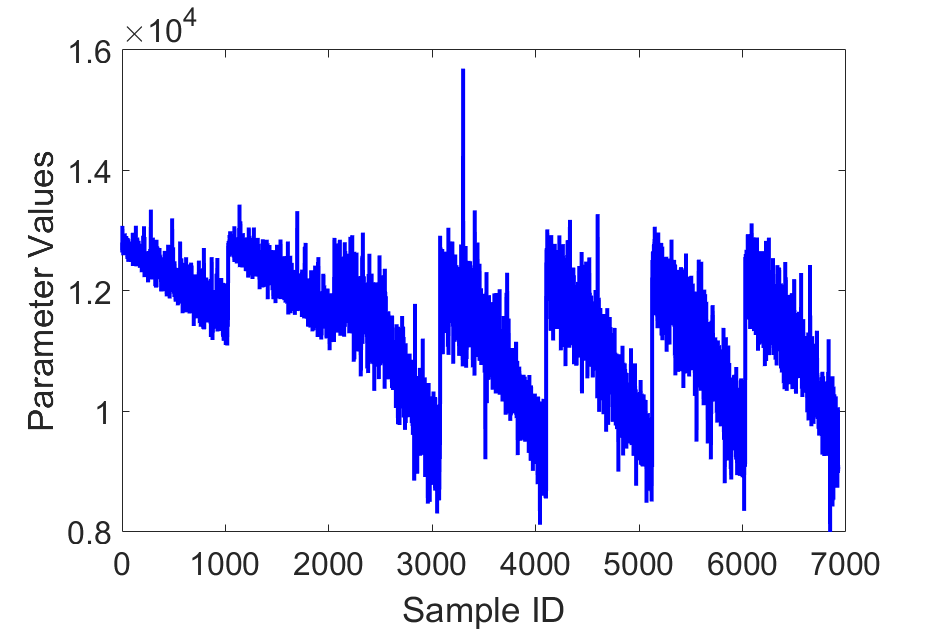}
}
\caption{Values of parameters in the cubic polynomial functions for all stress-strain curves considered for: (a) $a_1$; (b) $a_2$; (c) $a_3$.}
\label{img:cubic_param}
\end{figure}

Here $a_1,a_2,a_3$ are the coefficients of the cubic polynomial function. The obtained coefficient values are shown in Figure~\ref{img:cubic_param}. We notice that the cubic polynomial function representations have fitting mean squared errors between 0.41 to 0.92 for all models considered. As the stress value ranges from 0 to around 20MPa, we believe such a fitting error is acceptable. Further, the cubic polynomial function can give a good representation of the stress-strain relationship.

\subsubsection{Ogden function representation}

As the HDPE manifests a non-linear behavior under compression, the stress-strain curve can also be fitted using the Ogden function through the non-linear fitting\cite{prabhakar2022densification}. We consider the incompressible case as shown in Equation~\ref{eqn:ogden_func}. Detailed derivations of the expression can also be found in our prior work\cite{prabhakar2022densification}.

\begin{equation}
    P(\lambda)=\sum_{i=1}^n \frac{2\mu_i}{\alpha_i}(\lambda^{\alpha_i-1}-\lambda^{-\frac{1}{2}\alpha_i-1})
\label{eqn:ogden_func}
\end{equation}

\begin{figure}[h!]
\centering
\subfigure[]{
  \includegraphics[width=0.45\textwidth,height=0.35\textwidth]{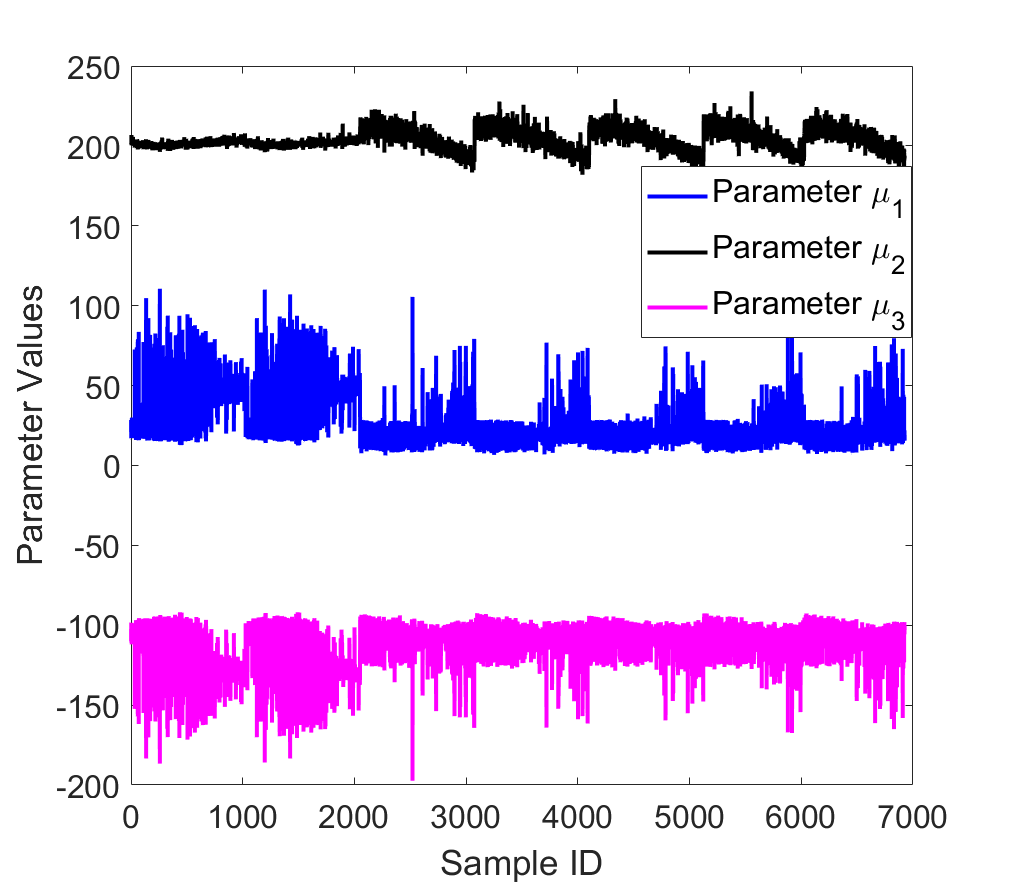}
}
\centering
\subfigure[]{
  \includegraphics[width=0.45\textwidth,height=0.35\textwidth]{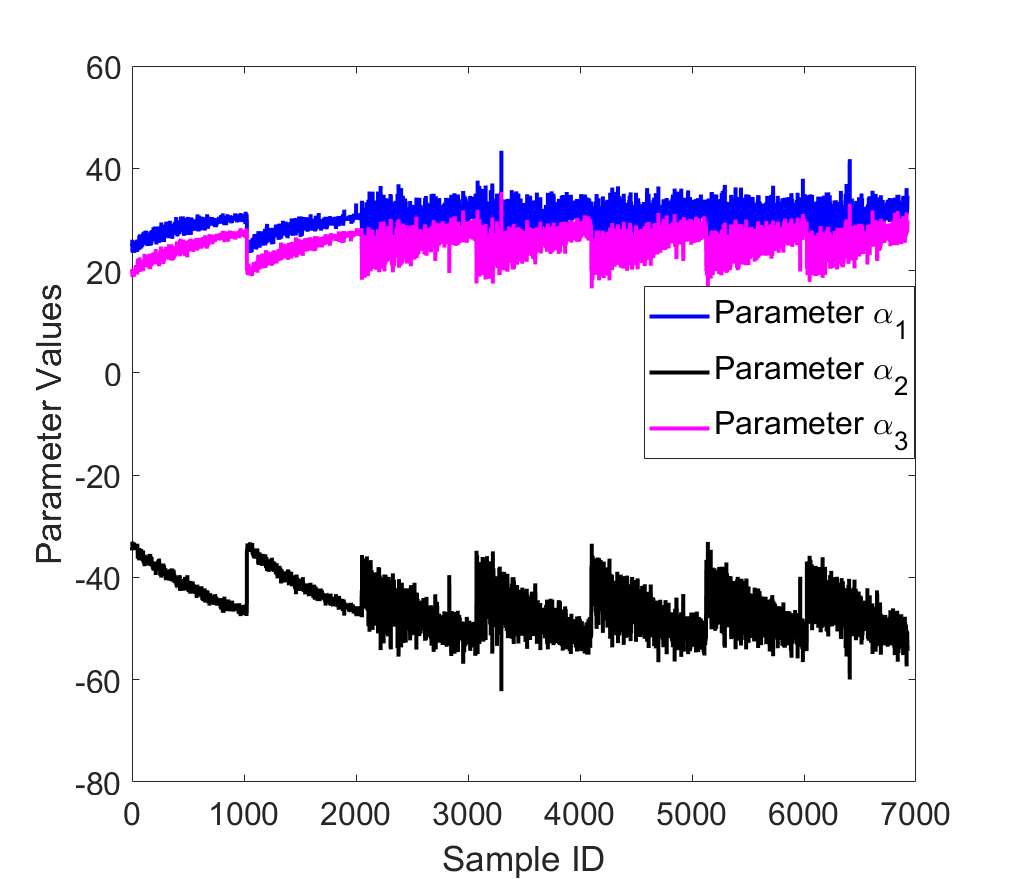}
}
\caption{Values of parameters in the Ogden functions for all stress-strain curves considered: (a) $\mu_1$, $\mu_2$ and $\mu_3$; (b) $\alpha_1$, $\alpha_2$ and $\alpha_3$.}
\label{img:ogden_param}
\end{figure}

This paper considers the 3rd-order Ogden model, which has $n=3$ in Equation~\ref{eqn:ogden_func}. Through non-linear regression, we can obtain the values of coefficients in the Ogden function as shown in Figure~\ref{img:ogden_param}. From non-linear regression, we notice that the Ogden function fitting errors are between 0.02 to 0.54. Since $\alpha_i$ appears on the denominator, we want to keep the sign of $\alpha_i$ consistent for all models in the non-linear regression such that the Neural Network is less likely to predict a denominator value close to zero. This issue is avoided by adding constraints to the parameter values during non-linear regression, such that the sign of the regression coefficients will always be positive or negative. From Figure~\ref{img:ogden_param}, the signs are consistent for all $\alpha_i$. Consequently, the Ogden function can also represent the stress-strain curve. 

\begin{figure}[h!]
\centering
	\includegraphics[width=0.6\textwidth]{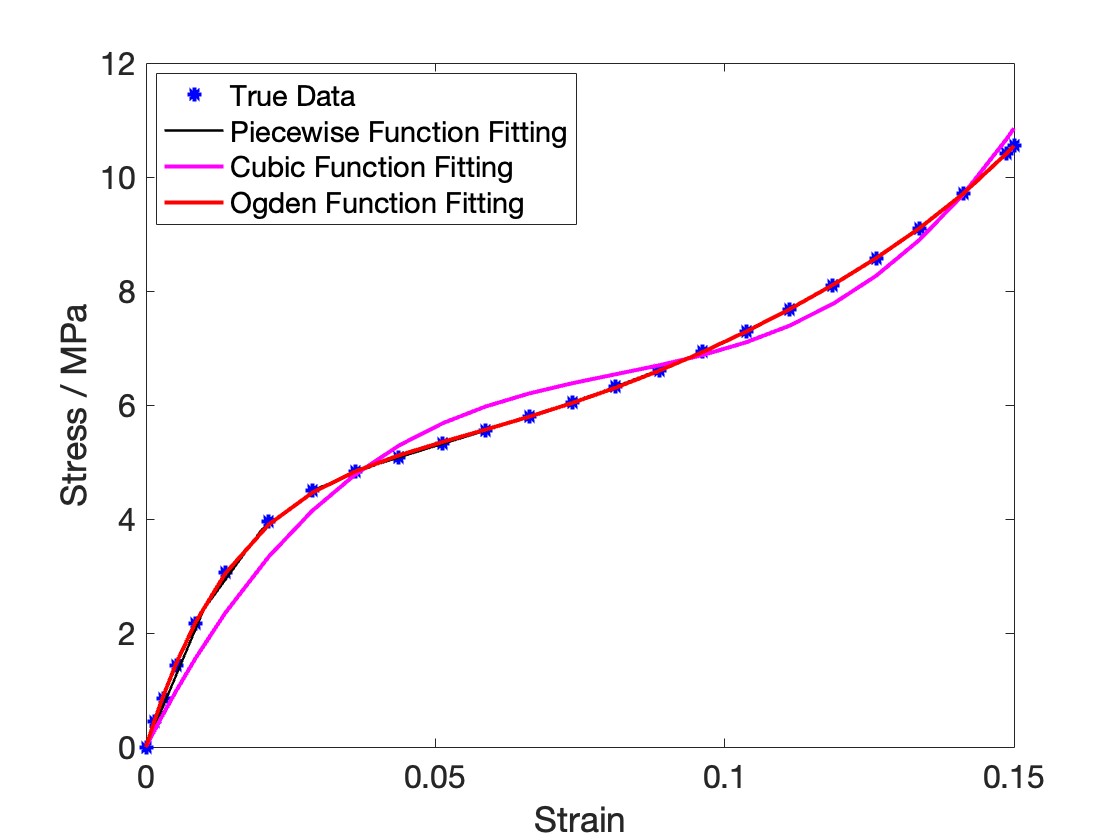}
	\caption{An example of stress-strain curve representation: blue dots are the true data, the black curve is fitted using linear piecewise function, the pink curve is fitted using the cubic polynomial function, and the red curve is fitted using the Ogden function.}
\label{img:vf_vs_t}
\end{figure}

Figure~\ref{img:vf_vs_t} shows how each representation compares with the true stress-strain data. Here we notice that the linear piecewise function (black curve) can accurately capture the trend of the stress-strain curve in a non-smooth manner. The cubic polynomial function (pink curve) gives a smooth representation of the curve but is less accurate than the linear piecewise function. The Ogden function (red curve) gives the best approximation of true stress-strain curve, also a visually smoother solution than the linear piecewise function. This is because we use the Ogden hyperelastic model to fit the matrix material description for the FEA simulations that are used to generate the input composite stress-strain responses for this paper.

\subsection{Loss Function and Training Process}

We propose different training processes for different curve representations. We utilize the Mean Squared Error (MSE) as the loss function for linear piecewise and cubic polynomial function representations. For both representations, The Neural Network uses MSE as a loss function for both representations and is trained for 500 epochs, each with 30 steps. We propose a modified MSE loss function `MMSE' for training the Ogden function representation.
We know from the Ogden function that $P'(\lambda=1)=\frac{dP}{d\lambda} \; at \; (\lambda=1) =\mu_1+\mu_2+\mu_3$. The term $P'(\lambda=1)$ represents the slope at the initial point on the stress-strain curve and is independent of the choice of $\alpha_i$. From our non-linear regression results in Figure~\ref{img:ogden_param}, the value of $\mu_i$ is more spread out than $\alpha_i$. Thus adding this constraint to the loss function is important to constrain the prediction of different $\mu_i$ and improve the prediction accuracy. Moreover, as the Modification function $g(\lambda)$ is a quadratic function of $\lambda-1$, adding the Modification function will not change the value of $P'(\lambda=1)$. 

The MMSE can be written as Equation~\ref{eqn:ogden_mse}.
\begin{equation}
    %MMSE=MSE(y,\Tilde{y})+\alpha\cdot MSE[(P'( \lambda=1)|y)-(P'(\lambda=1)|\Tilde{y})]
    MMSE=MSE(y,\Tilde{y})+\alpha\cdot MSE[(P'(\lambda=1)\mid y)-(P'(\lambda=1)\mid \Tilde{y})]
\label{eqn:ogden_mse}
\end{equation}

Where MSE is defined in Equation~\ref{eqn:mse_func}, $\alpha$ is a constant weight added to the loss term generated by initial slope $P'(\lambda=1)$. In this paper, we use $\alpha=0.01$. However, during the training process, we notice that directly training the Network with MMSE will lead to a blown-up prediction since the additional loss term creates more difficulty in finding the global minimum. So instead, we propose a `\textbf{hybrid}' training process: the Neural Network is trained for 500 epochs. For the first 300 epochs, we utilize MSE as the loss function. Then we utilize MMSE for the subsequent 100 epochs and then use MSE again for the last 100. During this training process, MMSE tries to change the gradient descent direction from MSE and help avoid falling into saddle points or local minima.

\section{Results and Discussion} \label{se:ML}

In this section, we evaluate the performance of our Feature Extraction and Curve Prediction modules on different stress-strain curve representations. The Feature Extraction module is validated by checking if the `Latent Vector' can effectively predict back to the original geometry. The Curve Prediction module is validated by checking how close the predicted curve is to the true curve. We randomly split the data into 60\% training, 20\% cross-validation, and 20\% testing for all datasets used.

\subsection{Feature Extraction Module Results}

\begin{figure}[h!]
%\centering
%\subfigure[]{
%  \includegraphics[width=0.45\textwidth]{./figures/real_model_1.png}
%}
%\centering
%\subfigure[]{
%  \includegraphics[width=0.45\textwidth]{./figures/pred_model_1.png}
%}
\centering
\subfigure[]{
  \includegraphics[width=0.35\textwidth]{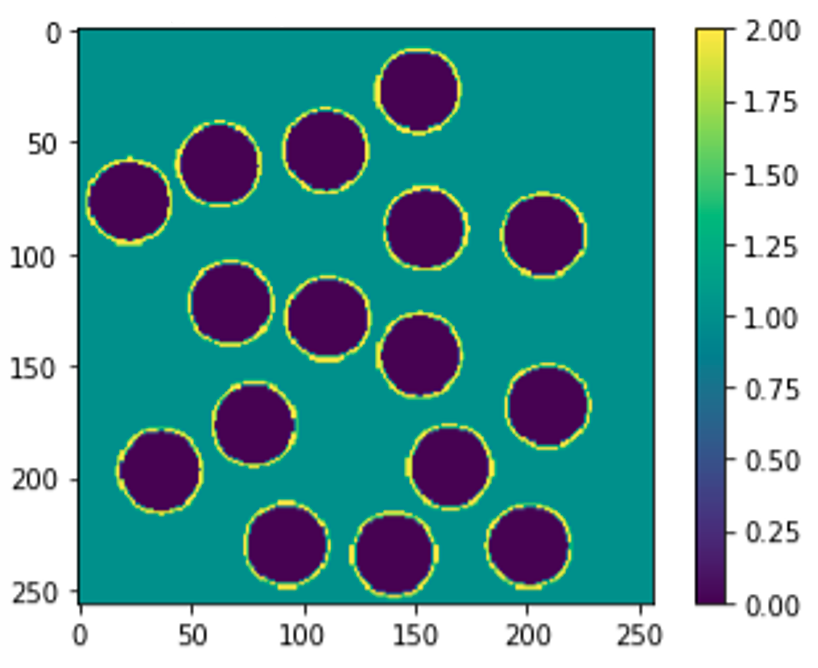}
}
\centering
\subfigure[]{
  \includegraphics[width=0.35\textwidth]{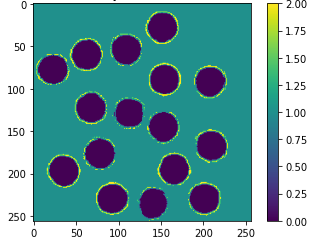}
}
\centering
\subfigure[]{
  \includegraphics[width=0.35\textwidth]{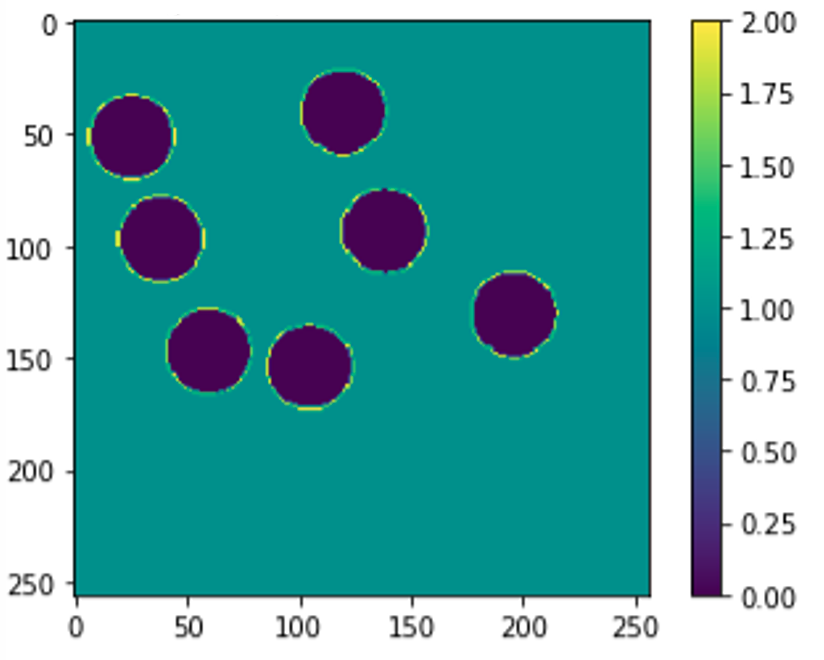}
}
\centering
\subfigure[]{
  \includegraphics[width=0.35\textwidth]{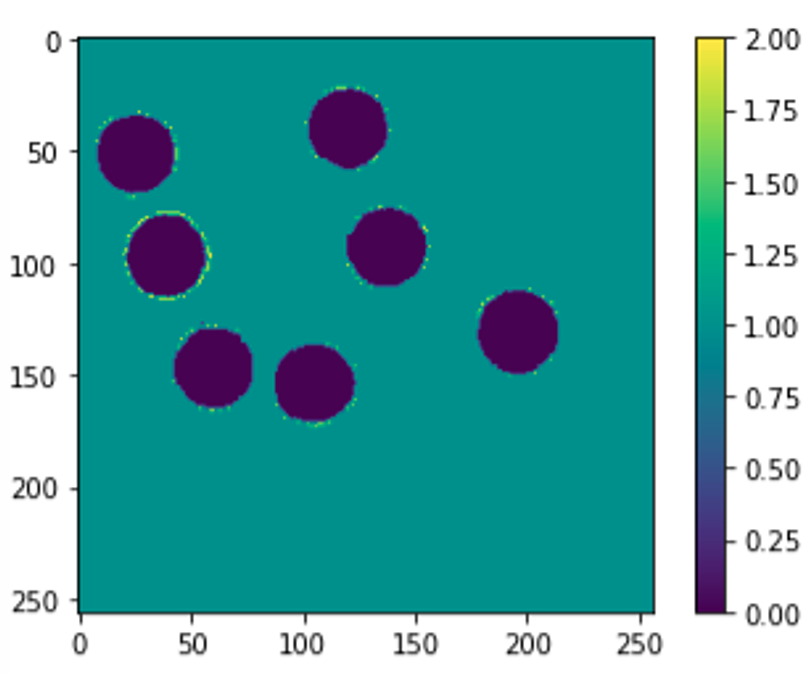}
}
\caption{True syntactic foam geometry versus Predicted syntactic foam geometry obtained from the Feature Extraction module. Thick-wall syntactic foam - (a) real and (b) predicted geometry. Thin-wall syntactic foam - (c) real and (d) predicted geometry}
\label{img:feature_extract_results}
\end{figure}

We first train the Feature Extraction module using 18000 syntactic foam geometries, with 60\% training, 20\% testing, and 20\% cross-validation. Figure~\ref{img:feature_extract_results} shows the predicted syntactic foam geometries from the Feature Extraction module. We can see that the predicted syntactic foam geometries can capture most of the critical features of the syntactic foams, even for complex geometries with high volume fractions, like Figure~\ref{img:feature_extract_results}(a) and (b). 

Moreover, to quantitatively measure how the Feature Extraction Module performs, we define an error rate (ER) to measure the prediction accuracy, as:

\begin{equation}
    ER = \frac{\sum_{i=1}^{N_x}\sum_{j=1}^{N_y} \mathbbm{1}_{[G_o(i,j) \neq G_p(i,j)]}}{N_x*N_y}
\end{equation}

where $N_x=N_y=256$ is the size of the geometry Cartesian Map, mentioned in \hbox{Section~\ref{sec:geom_represent}. $G_o$} represents the original geometry Cartesian Map, and $G_p$ represents the predicted geometry Cartesian Map. The error rate calculates the ratio between incorrectly predicted node values in the Cartesian Map and the total Cartesian Map nodes ($N_x*N_y$). To evaluate how the training dataset affects the performance of the Feature Extraction Module, we first pick 2000 samples as the test set, and train the module with different training data sizes. We use three random split methods (we call random split seed) and calculate the error rate with different random splits to account for randomness.

\begin{figure}[h!]
\centering
\subfigure[]{
  \includegraphics[width=0.3\textwidth]{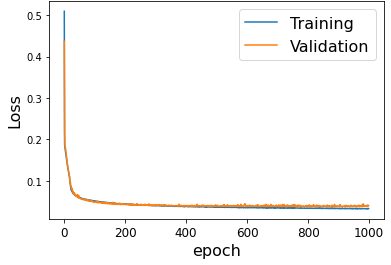}
}
\centering
\subfigure[]{
  \includegraphics[width=0.3\textwidth]{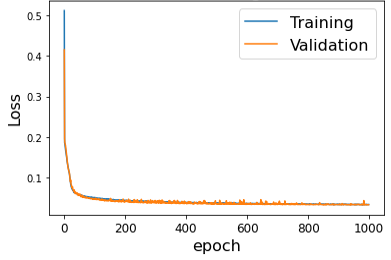}
}
\centering
\subfigure[]{
  \includegraphics[width=0.3\textwidth]{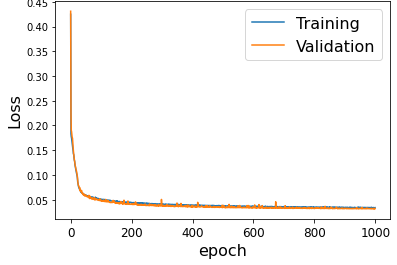}
}
\caption{Training loss profile of Feature Extraction Module (a) using 7000 samples (b) using 10000 samples (c) using 12500 samples}
\label{img:training_loss}
\end{figure}

\begin{figure}[h!]
\centering
\subfigure[]{
  \includegraphics[width=0.3\textwidth]{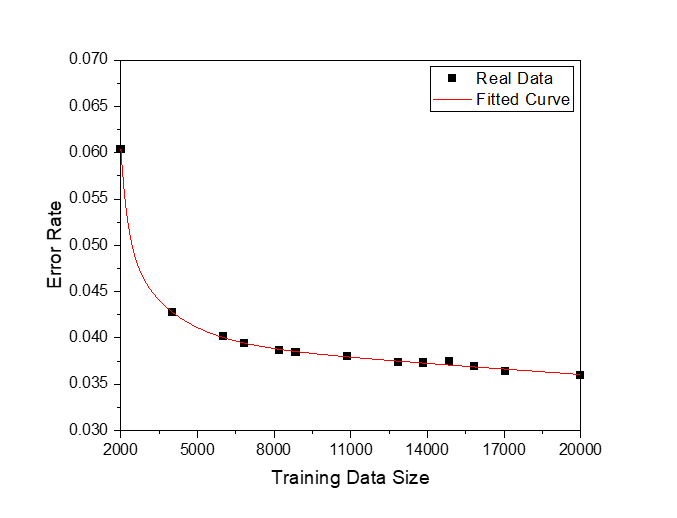}
}
\centering
\subfigure[]{
  \includegraphics[width=0.3\textwidth]{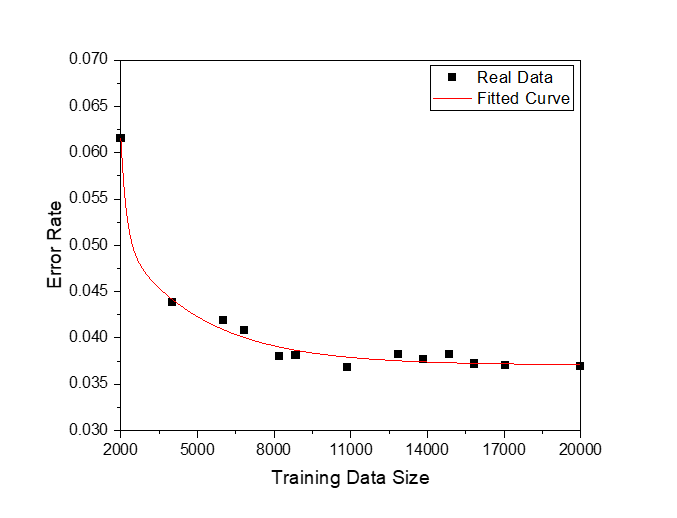}
}
\centering
\subfigure[]{
  \includegraphics[width=0.3\textwidth]{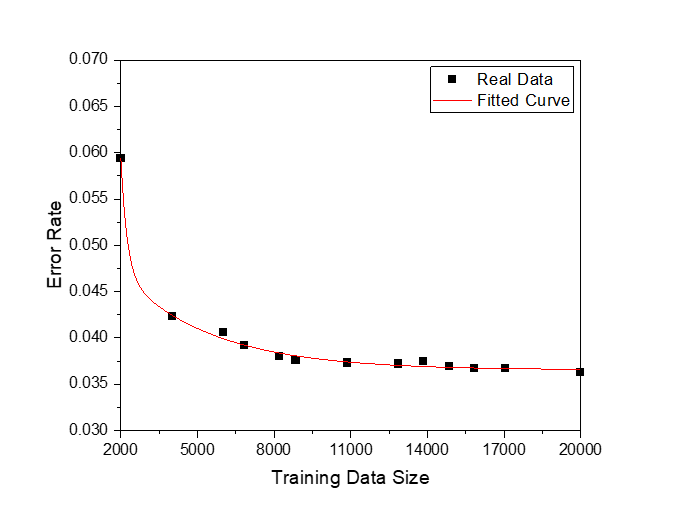}
}
\caption{Prediction error rate of Feature Extraction Module, on 2000 testing samples (a) random split seed 1 (b) random split seed 2 (c) random split seed 3}
\label{img:feature_extract_curve}
\end{figure}

\hbox{Figure~\ref{img:training_loss}} shows how the training loss decreases as we increase the training epochs (we run a total of 1000 epochs and each epoch has 40 steps) for training sample sizes of 7000, 10000, and 12500 each. We observe that the Feature Extraction Module optimization is converged within 1000 epochs. Further, \hbox{Figure~\ref{img:feature_extract_curve}} shows how the prediction error changes with increasing training data size. Despite different random split methods, the prediction error rate decreases as we increase the training sample size. When the training data size is larger than 15000, we have a consistent prediction error rate below 3.5\%. Thus in this paper, we choose a training data size of 18000 to train the Feature Extraction Module for extracting the corresponding high-level latent features of the syntactic foam geometries.

\begin{comment}
\begin{table}[h!]
\centering
\caption{Feature Extract Module prediction error}
\resizebox{0.8\textwidth}{!}{
\begin{tabular}{ccccc}
\hline
           & Random Split 1 & Random Split 2 & Random Split 3 & Average \\ \hline
Error Rate & 0.0353         & 0.0357         & 0.0349         & 0.0353  \\ \hline
\end{tabular}
}
\label{tab:error_feature_extraction}
\end{table}
\end{comment}

\subsection{Curve Prediction Module Results}

After fully training the Feature Extraction module, we can use the extracted feature vectors to predict the stress-strain curves using the Curve Prediction module. The prediction error is evaluated based on the errors at 21 uniformly distributed points along the curve (same as the discrete point locations when defining linear piecewise function). The errors are calculated based on Norm-2 Error (NE), defined in Equation~\ref{eqn:ne}.

\begin{equation}
    NE  = \sum_{i=1}^{n} (Y(i)-\hat{Y}(i))^2
\label{eqn:ne}
\end{equation}

\begin{figure}[h!]
\centering
\subfigure[]{
  \includegraphics[width=0.47\textwidth]{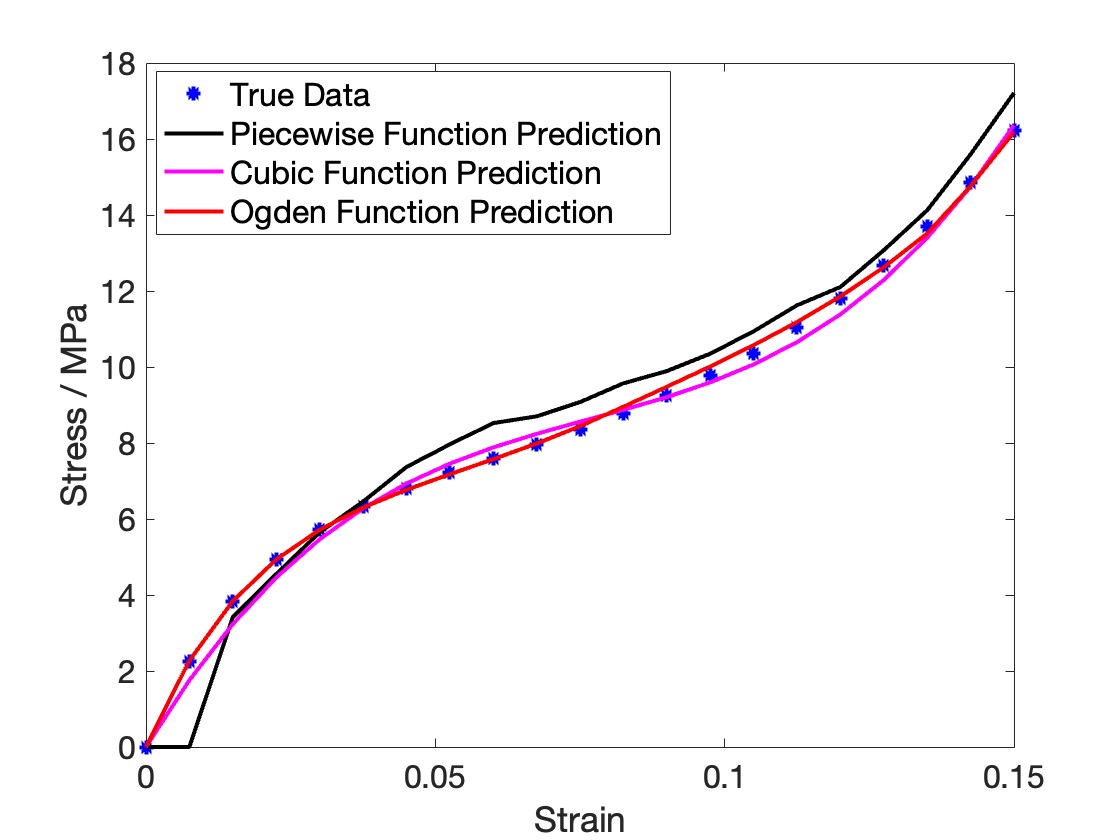}
}
\centering
\subfigure[]{
  \includegraphics[width=0.47\textwidth]{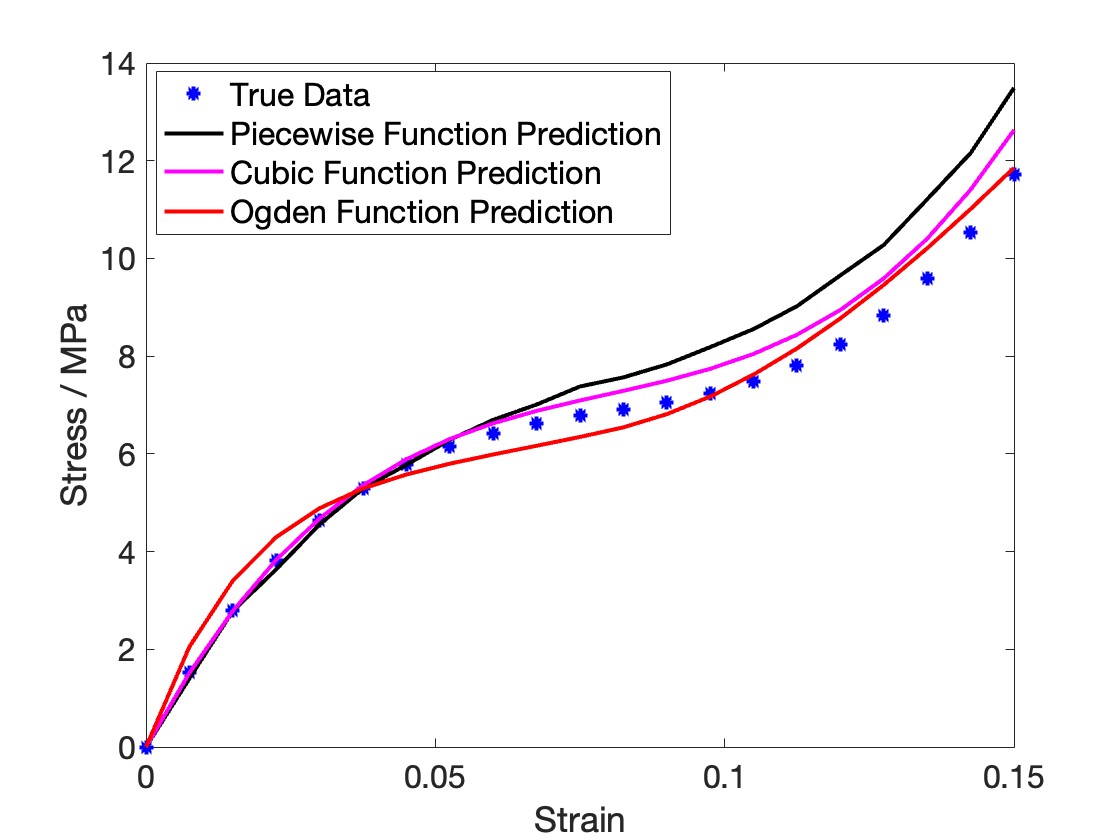}
}
\caption{Predicted stress-strain curves from the Curve Prediction module: (a) Example 1 (b) Example 2. The blue dots are the true data, the black curve is fitted using the linear piecewise function, the pink curve is fitted using the cubic polynomial function, and the red curve is fitted using the Ogden function.}
\label{img:self_supervise}
\end{figure}

$n=21$ represents the total size of the data points considered in one sample. The overall prediction errors are evaluated based on Mean Norm-2 Error and Max Norm-2 Error, which calculates the average and maximum Norm-2 Error in the testing sample set. Table~\ref{tab:curve_pred} shows the prediction errors of different stress-strain curve representation methods using PBNN. Besides, Figure~\ref{img:self_supervise} gives two examples of the Stress-Strain curve prediction using different representations.

\begin{table}[h!]
\centering
\caption{Curve Prediction Error for different curve representations with PBNN}
\resizebox{0.8\textwidth}{!}{
\begin{tabular}{ccc}
\hline
 & Mean Norm-2 Error & Max Norm-2 Error \\ \hline
Linear Piecewise Function    & 7.31         & 21.64              \\
Cubic Polynomial Function    & 5.03         & 18.73              \\
Ogden Function    & 5.89         & 14.89              \\ \hline
\end{tabular}
}
\label{tab:curve_pred}
\end{table}

Comparing the predictions for different representations, we can conclude that:
\begin{enumerate}
    \item The Ogden function and cubic polynomial function representations give similar prediction accuracy by providing a smooth curve close to the true curve. The polynomial function representation has a lower Mean Norm-2 Error, but Ogden function representation has a lower Max Norm-2 Error. This is because the cubic polynomial function has fewer parameters to train, making obtaining a better prediction during the training process easier. However, as shown in Figure~\ref{img:vf_vs_t}, the cubic polynomial function has fitting errors to the true curve.
    \item The linear piecewise function representation gives a relatively worst prediction by having the most significant prediction error. Moreover, the linear piecewise function gives a non-smooth prediction, and the relative relationship between adjacent points might not obey the physical truth (like the horizontal line at the beginning of the blue curve in Figure~\ref{img:self_supervise}(a), which is not true). Furthermore, we potentially introduce an additional error if we attempt to smooth the prediction obtained from the linear piecewise function. 
\end{enumerate}

\subsection{Importance of Feature Extraction module, Modification module, and Hybrid Training Process}

We introduce the ideas of different modules and training methods for PBNN. To understand the importance of the different aspects in PBNN, like the Feature Extraction module, Modification module, and the hybrid training process (only for Ogden function representation), this section compares PBNN's prediction error with several baseline models. We use the Ogden function representation for illustration purposes.

\begin{figure}[h!]
\centering
\subfigure[]{
  \includegraphics[width=0.45\textwidth]{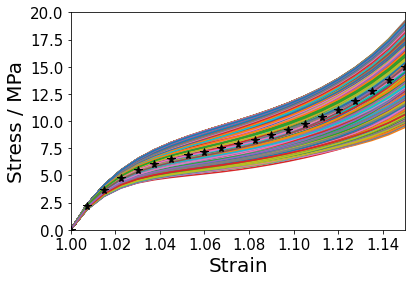}
}
\centering
\subfigure[]{
  \includegraphics[width=0.45\textwidth]{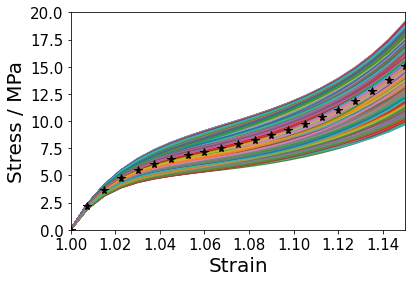}
}
\caption{Comparison between the mean curve of the Baseline model 0 (\textbf{black asterisk}) and (a) training dataset (b) testing dataset}
\label{img:baseline_0}
\end{figure}

\paragraph{Baseline model 0}
We first choose a baseline model that always outputs the mean stress-strain curve from the training dataset. The norm-2 error for the Baseline model 0 represents the variance of the testing dataset. To show the improved efficacy of other baseline models proposed, their norm-2 error should be lower than that of the Baseline model 0. Figure~\ref{img:baseline_0} shows a comparison between the mean curve and the training (a) and testing (b) datasets. 

\paragraph{Baseline model 1}
As discussed in Section~\ref{sec:feature_extraction}, we utilize the Feature Extraction module to extract latent space features using easily obtained syntactic foam geometries. To validate the effectiveness of the Feature Extraction module, we consider a baseline model in which we directly use an Encoder structure (as discussed in Figure~\ref{img:Feature_Extraction}) to predict the Ogden parameters (named Baseline-1).

\paragraph{Baseline model 2}

\begin{figure}[h!]
\centering
  \includegraphics[width=0.95\textwidth]{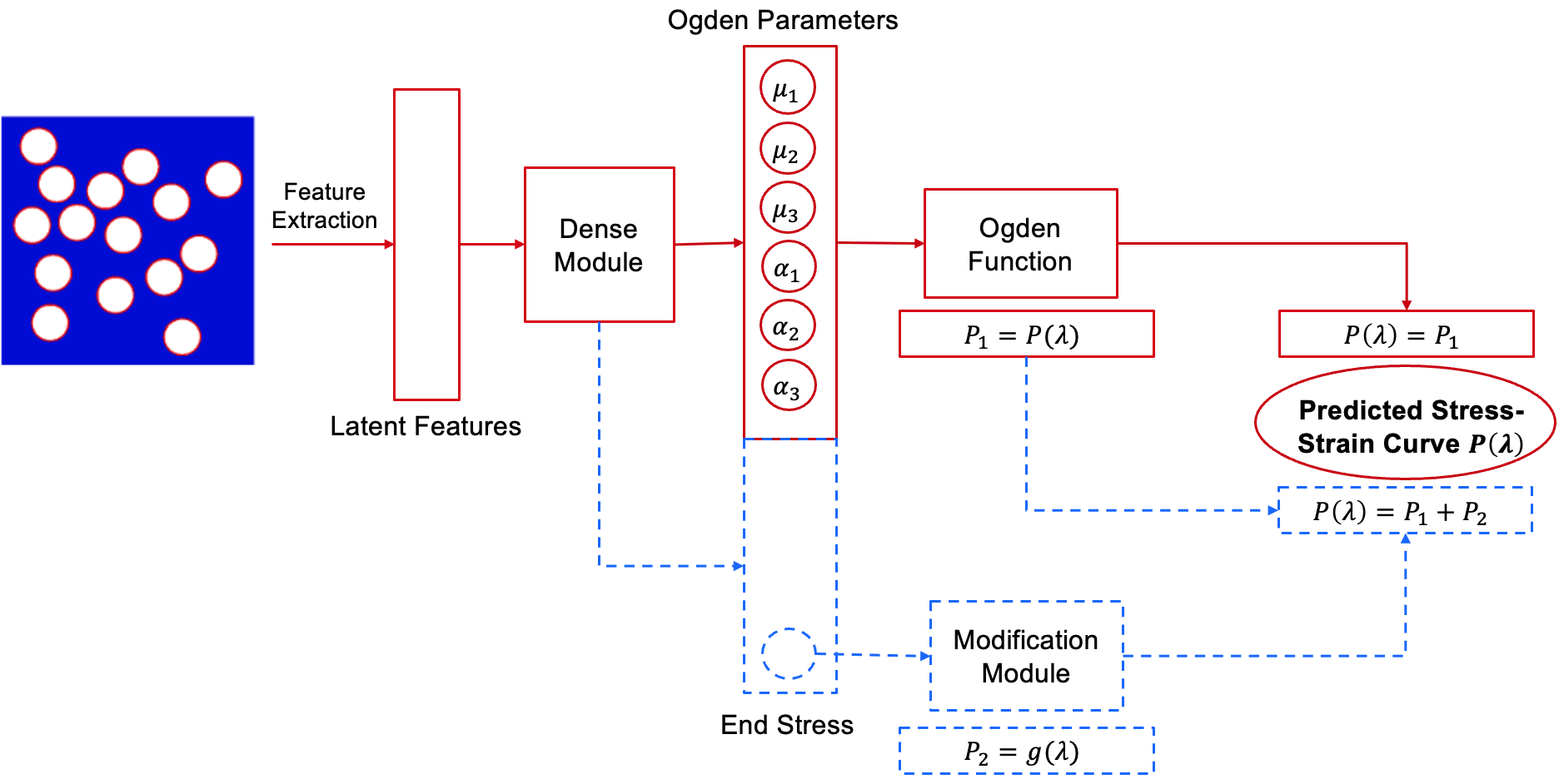}
\caption{Stress-Strain curve prediction using PBNN framework without modification module (Baseline-2) is shown using red color. PBNN with modification module is shown by adding blue dashed line modules to Baseline-2. Baseline-2 directly predicts the Ogden parameters from extracted latent features and constructs the Stress-Strain curve with the Ogden function. While PBNN also uses the Modification module to generate the Modification function $P_2$ and constructs the Stress-Strain curve by adding the Ogden function and Modification function.}
\label{img:pbnn_mod}
\end{figure}

To validate how the Modification module performs, we choose another baseline model by removing the Modification module from PBNN (named Baseline-2). We compare the prediction between PBNN with and without the Modification module. The detailed frameworks for the two models are shown in Figure~\ref{img:pbnn_mod}, with and without the blue dashed line modules.

\paragraph{Baseline model 3 (only for Ogden function representation)}

To understand how our proposed hybrid training improves the prediction accuracy, we consider another baseline model (named Baseline-3), which has the same structure as PBNN but only uses MSE as the loss function instead of MMSE.

\paragraph{Prediction results}

Here we show the prediction accuracy using our training data with 6825 syntactic foam models. Three different data-splitting methods are used, and the average error is calculated. The results are shown in Table~\ref{tab:error_baseline}. From the results, we notice that:
\begin{enumerate}
    \item The Baseline-0 has the largest mean Norm-2 Error compared to all the other models, while the max norm-2 error is lower than Baseline-1 and Baseline-2. This means Baseline-1 and Baseline-2 could make predictions beyond the range of the test set curves. On the other hand, Baseline-3 and PBNN have better predictions than Baseline-0 for both mean norm-2 error and max norm-2 error.
    \item Baseline-2 has slightly better prediction than Baseline-1, meaning that the Feature Extraction module can extract the high-level features better and improve the curve prediction accuracy.
    \item Baseline-3 significantly improves the prediction accuracy compared to Baseline-1 and Baseline-2, proving that the modification module is essential to achieving better prediction accuracy.
    \item PBNN can give a better prediction than Baseline-3, proving that the hybrid training procedure improves the prediction accuracy.
\end{enumerate}

\begin{table}[h!]
\caption{Curve Prediction Error for different frameworks}
\centering
\resizebox{0.7\textwidth}{!}{
\begin{tabular}{ccc}
\hline
           & Mean Norm-2 Error & Max Norm-2 Error \\ \hline
Baseline-0 & 8.14              & 18.26             \\
Baseline-1 & 7.51              & 30.12              \\ 
Baseline-2 & 6.69              & 30.08              \\ 
Baseline-3 & 6.14              & 18.18              \\ 
PBNN       & 5.89              & 14.89              \\ \hline
\end{tabular}
}
\label{tab:error_baseline}
\end{table}

\begin{figure}[h!]
\centering
\subfigure[]{
  \includegraphics[width=0.4\textwidth]{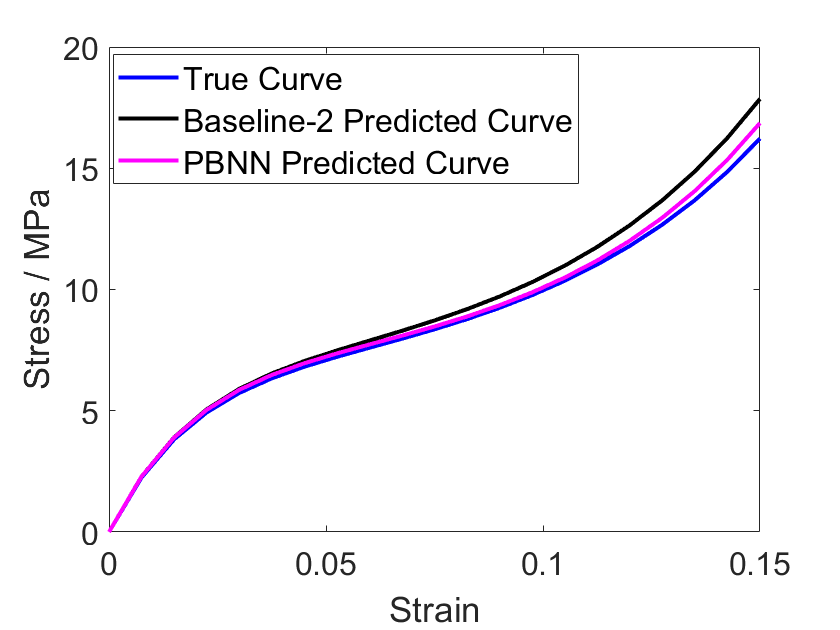}
}
\centering
\subfigure[]{
  \includegraphics[width=0.4\textwidth]{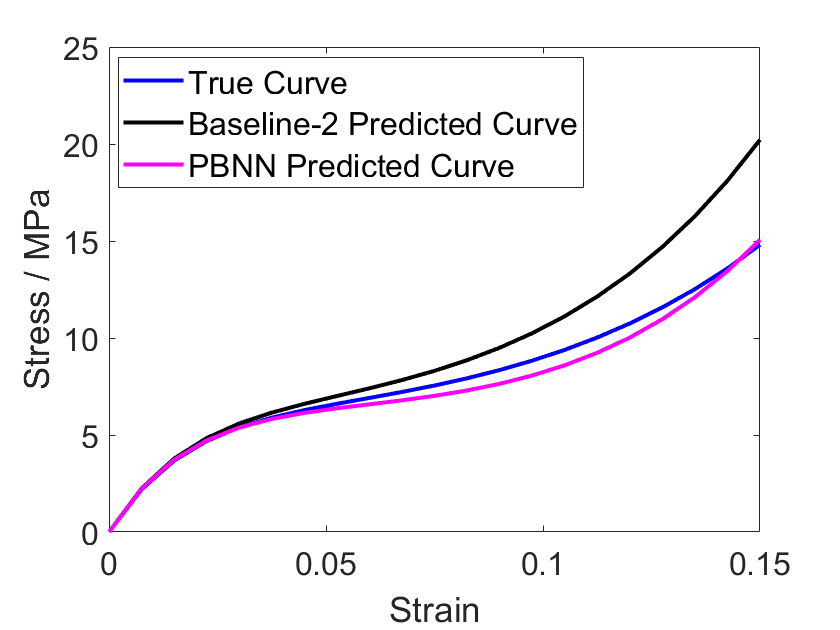}
}
\caption{Importance of the modification module: stress-strain curve prediction between Baseline-2 and PBNN on test set samples: (a) Curve prediction on test sample 1 (b) Curve prediction on test sample 2. The blue curve is the true Stress-Strain curve, the green curve is the predicted curve from Baseline-2, and the red curve is the predicted curve from PBNN.}
\label{img:pbnn_curve}
\end{figure}

Moreover, we can visualize the effect of the modification module in Figure~\ref{img:pbnn_curve}, which validates the predictions of PBNN and Baseline-2 for two randomly picked test datasets (test set 1 and test set 2) from the total 20\% test data set. The graph shows that the modification module could effectively push the initially predicted curve closer to the true curve and enhance the stress-strain curve prediction accuracy.

\begin{comment}
    
\subsection{Summary of the Results}

The above results and conclusions can be summarized as the following:

\begin{enumerate}
    \item Comparing to traditional Piecewise Function representation, our proposed Cubic, and Ogden function representations can reduce the prediction error and have similar prediction performances. The Ogden function can better match the true stress-strain data and provides a more 'physically meaningful' representation. In contrast, the Cubic function is easier to train as it has a relatively simple expression and fewer parameters. 
    \item We can further extend the Cubic function representation to other different curve prediction problems as the function is generated using a general polynomial function.
    \item By comparison with several baseline models, we prove our proposed PBNN framework can give the best stress-strain curve prediction, and the modules inside PBNN are necessary for better prediction accuracy.
\end{enumerate}

\end{comment}

\section{Conclusions}

This paper proposes a Parameterization-base Neural Network (PBNN) framework to predict the non-linear stress-strain responses of composites. We choose syntactic foam composites to develop these frameworks due to their complex internal architecture and corresponding mechanical responses. Instead of predicting discrete points along the stress-strain curve, we propose the stress-strain curve representations using the cubic polynomial function and Ogden function and utilize PBNN to predict corresponding function parameters. This vastly reduces the computational cost and data size needed for training. By comparing different baseline models, we further show that PBNN achieves a better prediction accuracy of the stress-strain curve than other baseline models.

The main conclusions and contributions of this paper are:
\begin{enumerate}
    \item This is the first attempt to our knowledge to predict non-linear stress-strain responses by treating them as a parameterized function, especially for the complex composite material analysis. 
    \item We have shown that our method could simplify the Machine Learning problem and generate a `physically meaningful' prediction by utilizing the Feature Extraction module and the Modification module. The Feature Extraction module extracts the high-level features from the microstructure geometry into a latent vector, serving as reduced-order input to the ML framework. The Modification module improves the prediction accuracy by referring to an auxiliary prediction and reconstructing the predicted stress-strain curve expression. 
    
    \item We have demonstrated that PBNN can predict general polynomial functions (like a cubic polynomial function) or complex highly non-linear functions (like an Ogden function) from internal material microstructures.
    
    \item Our method is not limited to syntactic foam or composite material stress-strain prediction, and we can use a similar approach for all curve-related predictions. Our method can also be extended to develop knowledge/physics-guided Machine Learning algorithms with the proposed Feature Extraction module and Modification module.
\end{enumerate}

\section*{Acknowledgements}
The authors would also like to acknowledge the support from the University of Wisconsin Graduate Fellowship for partially supporting Haotian Feng's doctoral studies.

%\section*{Author Contributions}
%P.P. and H.F. conceptualized and developed this study. H.F. implemented the research presented in this paper. H.F. and P.P. evaluated the outcomes of the work and drafted the manuscript. 

\section*{Declarations}

%\textcolor{red}{Some journals require declarations to be submitted in a standardised format. Please check the Instructions for Authors of the journal to which you are submitting to see if you need to complete this section. If yes, your manuscript must contain the following sections under the heading `Declarations':}

\begin{itemize}
\item Funding \\
U.S. Department of Defense (DoD) Office of Naval Research - Young Investigator Program (ONR - YIP) Grant [N00014-19-1-2206]: {\em{Sea-based Aviation: Structures and Materials Program}}. 
\\

\item Conflict of interest/Competing interests \\
%(check journal-specific guidelines for which heading to use)
Not applicable
\\

\item Ethics approval \\
Not applicable
\\

\item Consent to participate\\
Not applicable
\\

\item Consent for publication\\
Not applicable
\\

%\item Availability of data and materials\\
%Data will be made available by the authors upon reasonable request.
%\\

\item Code and Data availability \\
The entire PBNN framework and baseline models can be found on our GitHub page: \url{https://github.com/Isaac0047/Parameterization-based-Neural-Network.git}. This includes the entire implementation code with model generation, data processing, PBNN framework setup and baseline models setup. A detailed description the steps involved with running the PBNN framework is included in the readme file on this GitHub page. 
\\

\item Authors' contributions \\
P.P. and H.F. conceptualized and developed this study. H.F. implemented the research presented in this paper. H.F. and P.P. evaluated the outcomes of the work and drafted the manuscript. 

\end{itemize}

%\textcolor{red}{If any of the sections are not relevant to your manuscript, please include the heading and write `Not applicable' for that section. }

%\section*{Data Availability}\label{dataAvail}
%Data will be made available by the authors upon reasonable request.


\section*{Acknowledgements}
The authors would like to acknowledge the support from the U.S. Department of Defense (DoD) Office of Naval Research - Young Investigator Program (ONR - YIP) Grant [N00014-19-1-2206] through the {\em{Sea-based Aviation: Structures and Materials Program}}. The authors would also like to acknowledge the support from the University of Wisconsin Graduate Fellowship for partially supporting Haotian Feng's doctoral studies.

%\section*{Author Contributions}
%P.P. and H.F. conceptualized and developed this study. H.F. implemented the research presented in this paper. H.F. and P.P. evaluated the outcomes of the work and drafted the manuscript. 

\section*{Data Availability}\label{dataAvail}
Data will be made available by the authors upon reasonable request.

%%%%%%%%%%%%%%%%%%%%%%%%%%%%%%%%%%%%%%%%%%

%\section*{References}
{\footnotesize
%\singlespacing
\bibliographystyle{unsrt}
%\biboptions{sort&compress}
\bibliography{sn-bibliography}
}
%----------------------------------------------------------------------------------------

\end{document}